\documentclass[12pt]{article}
\usepackage{amsmath}
\usepackage{graphicx,psfrag,epsf}
\usepackage{enumerate}
\usepackage{url} 

\newcommand{\blind}{0}

\allowdisplaybreaks

\usepackage{amsthm,amsfonts}
\usepackage{comment}
\usepackage{color}
\usepackage[bf]{caption}
\usepackage{lscape}
\usepackage[backend=bibtex, style=nature, citestyle=authoryear]{biblatex}
\bibliography{bibliography}

\def\inprob{\stackrel{p}{\rightarrow}}
\def\indist{\rightsquigarrow}
\def\ind{\perp\!\!\!\perp}
\def\T{{ \mathrm{\scriptscriptstyle T} }}

\newcommand{\Pb}{\mathbb{P}}
\newcommand{\Pn}{\mathbb{P}_n}

\newcommand{\E}{\mathbb{E}}
\newcommand{\R}{\mathbb{R}}

\newcommand{\bO}{\mathbf{O}}
\newcommand{\bo}{\mathbf{o}}
\newcommand{\bX}{\mathbf{X}}
\newcommand{\bx}{\mathbf{x}}
\newcommand{\bV}{\mathbf{V}}
\newcommand{\bv}{\mathbf{v}}
\newcommand{\bW}{\mathbf{W}}
\newcommand{\bw}{\mathbf{w}}
\newcommand{\bS}{\mathbf{S}}
\newcommand{\bs}{\mathbf{s}}

\newcommand{\bg}{\mathbf{g}}
\newcommand{\bh}{\mathbf{h}}

\newcommand{\bpsi}{\boldsymbol\psi}

\def\supp{\text{supp}}
\DeclareMathOperator*{\argmin}{arg\,min}

\DeclareSymbolFont{bbold}{U}{bbold}{m}{n}
\DeclareSymbolFontAlphabet{\mathbbold}{bbold}
\newcommand{\one}{\mathbbold{1}}
\newtheorem{theorem}{Theorem}

\theoremstyle{remark}
\newtheorem{assumption}{Assumption}
\newtheorem{assumptionprime}{Assumption}

\newtheorem{remark}{Remark}

\begin{document}

\def\spacingset#1{\renewcommand{\baselinestretch}%
{#1}\small\normalsize} \spacingset{1}


\if0\blind
{
  \title{ \vspace{-.3in} \bf Robust causal inference with \\ continuous instruments using the \\ local instrumental variable curve}
  \author{Edward H. Kennedy
    \thanks{Edward Kennedy is Assistant Professor of Statistics, Carnegie Mellon University, Pittsburgh, PA 15217 (e-mail: edward@stat.cmu.edu). Scott Lorch is Associate Professor of Pediatrics at the University of Pennsylvania School of Medicine and Attending Neonatologist at The Children's Hospital of Philadelphia. Dylan Small is Professor of Statistics, University of Pennsylvania. The authors gratefully acknowledge support from NIH grant R01-DK090385 (Kennedy) and NSF grant SES-1260782 (Small).}\hspace{.2cm}
    \\   \date{}
    Department of Statistics, Carnegie Mellon University \\ \\
    Scott A. Lorch \\
    Department of Pediatrics, Perelman School of Medicine, \\ University of Pennsylvania \\ \\
    Dylan S. Small \\
    Department of Statistics, The Wharton School, University of Pennsylvania }
  \maketitle
  \setcounter{page}{0}
  \thispagestyle{empty}

} \fi

\if1\blind
{
  \bigskip
  \bigskip
  \bigskip
  \begin{center}
    {\LARGE\bf Robust causal inference with \\ \vspace{.15in} continuous  instruments  using the \\ \vspace{.25in}  local instrumental variable curve}
  \end{center}
  \setcounter{page}{0}
  \medskip
} \fi

\vspace{-.1in}
\begin{abstract}
Instrumental variables are commonly used to estimate effects of a treatment afflicted by unmeasured confounding, and in practice instruments are often continuous (e.g., measures of distance, or treatment preference). However, available methods for continuous instruments have important limitations: they either require restrictive parametric assumptions for identification, or else rely on modeling both the outcome and treatment process well (and require modeling effect modification by all adjustment covariates). \textcolor{black}{In this work we develop the first semiparametric doubly robust  estimators of the local instrumental variable effect curve}, i.e., the effect among those who would take treatment for instrument values above some threshold and not below. In addition to being robust to misspecification of either the instrument or treatment/outcome processes, our approach also incorporates information about the instrument mechanism and allows for flexible data-adaptive estimation of effect modification. We discuss asymptotic properties under weak conditions, and use the methods to study infant mortality effects of neonatal intensive care units with high versus low technical capacity, using travel time as an instrument. 
\end{abstract}

\noindent%
\vfill

\thispagestyle{empty}

\newpage

\spacingset{1.45} 
\spacingset{1}

\section{Introduction}
\label{sec:intro}

Instrumental variables (IVs) are commonly used to estimate effects of treatments that are afflicted by unmeasured confounding. Instruments are special variables that influence treatment, but are themselves unconfounded and do not directly affect outcomes, allowing the recovery of some causal information from data that might otherwise be unusable. In practice, instruments are often continuous (e.g., measures of distance, or treatment preference), but most available methods only consider instruments that are discrete (and typically binary). Further, methods that do allow for continuous IVs have important limitations.

Classical IV methods (e.g., standard two-stage least squares), which were developed in a structural equation model framework, allow for continuous instruments but require strong parametric assumptions for identification, assume that treatment effects do not vary across units, and also require correct parametric models for at least how the outcome process depends on covariates and instruments \autocite{wooldridge2010econometric, okui2012doubly}. Alternatively, Robins and others  \autocite{robins1989analysis, robins1994correcting, hernan2006instruments, tan2010marginal, okui2012doubly} developed approaches in the potential outcomes framework that can also handle continuous instruments, but which allow heterogeneous treatment effects, and also permit doubly robust covariate adjustment \autocite{robins2001inference,van2003unified,bang2005doubly}. Doubly robust IV methods are consistent as long as either the instrument mechanism or the treatment/outcome mechanisms are correctly modeled (not necessarily both), and they can also yield fast root-n convergence rates and inference even when using flexible nonparametric methods for covariate adjustment. However, the methods developed in this framework still require parametric assumptions for identification; they typically target treatment effects on the treated, and achieve identification with dimension-reducing parametric assumptions that restrict how heterogeneous treatment effects can be. As noted for example by \textcite{tchetgen2013alternative}, this kind of approach is problematic because a priori information about the parametric form of underlying causal structure is rarely available, and misspecification could lead to large biases that cannot be detected with data.

An alternative approach is to replace dimension-reducing homogeneity assumptions with a monotonicity assumption  \autocite{robins1989analysis, imbens1994identification}, which rules out the possibility that any units would respond oppositely to encouragement from the instrument. In other words, there can be units who are encouraged by the instrument, as well as units who do not respond at all to the instrument, but there cannot be units who defy encouragement from the instrument. For example, in the binary instrument case, there can be units who take treatment if and only if they receive the instrument, as well as units who always or never take treatment; however, there cannot be units who take control if the instrument is received but take treatment if not.
This assumption is often plausible in practice, and also permits nonparametric identification of causal effects among compliers (i.e., those who do respond to encouragement from the instrument). However, monotonicity is usually framed in terms of binary instruments \autocite{imbens1994identification, abadie2003semiparametric, tan2006regression, ogburn2015doubly}. An important exception is a strand of work that has focused on estimating local IV (LIV) curves, i.e., effects among units who would comply right at a given threshold value of the instrument \autocite{heckman1997instrumental, heckman1999local,glickman2000derivation, heckman2005structural}.

This literature on LIV approaches arose out of a latent index or selection model framework \autocite{vytlacil2002independence}, and is unique in allowing for continuous instruments while still permitting nonparametric identification. However there are important limitations. {First, available approaches for estimating the LIV curve rely on modeling how both the treatment and outcome depend on covariates and instrument \autocite{basu2007use, carneiro2010estimating}, and typically use restrictive parametric models. This is problematic since parametric models are often relied upon based on convenience, rather than real substantive knowledge, and can yield severe bias if misspecified. Conversely, fully nonparametric approaches are sensitive to the curse of dimensionality and typically yield estimators with slow rates of convergence, as well as little hope for centered confidence intervals without impractical undersmoothing {(we refer to Section 5.7 of \textcite{wasserman2006all} for details)}. Further, in the IV setting there may be some information available about how the instrument depends on covariates (e.g., about the density of the instrument given covariates), but this is not incorporated in approaches that rely solely on treatment and outcome models, whether parametric or nonparametric. Our paper solves these problems with a semiparametric doubly robust approach that can attain parametric rates of convergence, even while allowing flexible nonparametric estimation of nuisance functions.}

Second, available LIV estimands are fully conditional on all measured covariates, even though in many cases effect modification is not of particular scientific interest, or else it is only of interest for a small subset of covariates. Marginal effects {can often be estimated more robustly at faster rates of convergence,} and are often more closely tied to scientific questions. \textcite{van2003unified} point out that using fully conditional effects puts us at the whim of whatever confounders happen to arise in the dataset at hand, whereas marginal effects allow framing scientific questions a priori. Marginalizing currently available fully conditional estimators leads to awkward and uninterpretable models, as discussed in the local average treatment effect (LATE) setting by \textcite{ogburn2015doubly}. {However, in contrast to the LATE setting, currently available methods cannot be adapted for doubly robust estimation of the LIV curve, even in the simpler case where effect modifiers are discrete or not present. Our paper solves these problems by developing methods for working models of a marginal version of the LIV curve itself, allowing for arbitrary effect modification.  Our direct modeling approach eases interpretability by allowing analysts to incorporate background knowledge on the actual parameter of interest, without having to specify models for nuisance quantities that are not of direct scientific interest.}

In addition to the above, our work makes several other important advances. {Importantly, we use empirical process theory and sample splitting to derive asymptotic properties of our approach under weak conditions, which allow for flexible data-adaptive estimation of nuisance functions in the presence of complex high-dimensional confounding. We also develop a doubly robust cross-validation approach for model selection in high-dimensional settings, which is crucial for learning the LIV curve from data.} Finally, we explore finite-sample properties via simulation, and implement our methods to study effects of high-level neonatal intensive care units (NICUs) on infant mortality, using travel time as an instrument.

\section{Preliminaries}

\subsection{Data \& Notation}
\label{4data}

\begin{figure}[h!]
\begin{center}
\includegraphics[width=3.5in]{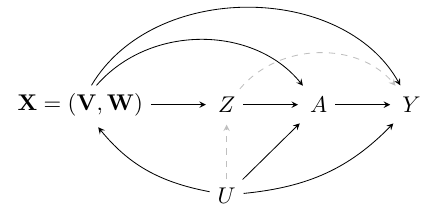}
\end{center}
\caption{Directed acyclic graph showing covariates $\bX$ (consisting of potential effect modifiers of interest $\bV$ and other variables $\bW$), instrument $Z$, treatment $A$, outcome $Y$, and unmeasured variables $U$. Gray dotted arrows indicate relationships that will be assumed absent by identifying assumptions. \label{fig:dag}}
\end{figure}

Suppose we observe an independent and identically distributed sample $(\bO_1,...,\bO_n)$ with $\bO=(\bX,Z,A,Y)$, where $\bX$ is a vector of covariates, $Z$ is a continuous instrument for a binary treatment $A$, and $Y$ is some real-valued outcome of interest. The covariates $\bX=(\bV,\bW)$ are partitioned into potential effect modifiers of interest $\bV$ and other covariates $\bW=\bX \setminus \bV$ not of interest but for which adjustment is still necessary. The choice of $\bV$ is based purely on the scientific question, so that if effect modification is not of interest one can simply select $\bV=\emptyset$. We characterize causal effects using potential outcome notation \autocite{rubin1974estimating}, and so let $Y^a$ (and $Y^{za}$) denote the potential outcomes that would have been observed had treatment level $A=a$ (and instrument level $Z=z$) been received. Similarly we let $A^z$ denote the potential treatment that would have been observed under instrument level $Z=z$. A directed acyclic graph showing the data structure is given in Figure~\ref{fig:dag}. 

We let $P$ denote the distribution of $\bO$, with density with respect to some dominating measure given by $p(\bo)=p(y \mid \bx,z,a) p(a \mid \bx,z) p(z \mid \bx) p(\bx)$. In general we write the density of a variable $T$ under $P$ evaluated at $z$ as $p(T=z)$, except when there is no ambiguity (e.g., $p(t)$ is the density of $T$ at $t$), and we use $\supp(T)$ for the support of $T$. Finally we use some additional notation to simplify the presentation. Specifically we let $\pi(z \mid \bx)=p(Z=z \mid \bX=\bx)$ denote the density of the instrument given covariates (i.e., instrument propensity score), $\mu(\bx,z)=\E(Y \mid \bX=\bx, Z=z)$ and $\lambda(\bx,z)=\E(A \mid \bX=\bx, Z=z)$ denote the outcome and treatment regression functions, respectively, with marginalized versions given by $m(z,\bv)=\E\{ \mu(\bX,z) \mid \bV=\bv\}$ and $\ell(z,\bv)= \E\{ \lambda(\bX,z) \mid \bV=\bv\}$. We let $\Pn$ denote the empirical measure so that empirical averages can be written as $n^{-1} \sum_i f(\bO_i)=\Pn\{f(\bO)\}$. The notation $|| \cdot ||$ denotes the Euclidean norm $|| \boldsymbol\beta ||=(\boldsymbol\beta^\T \boldsymbol\beta )^{1/2}$, and $||f||_2=\{\int f(\bo)^2 \ dP(\bo)\}^{1/2}$ denotes the $L_2(P)$ norm .

\subsection{Monotonicity}

Before defining the causal estimand of interest and considering identifying assumptions, it will be helpful to discuss the concept of monotonicity, which was introduced in various forms by \textcite{robins1989analysis} and \textcite{imbens1994identification}, among others. In a classical binary instrument setting (where supp$(Z)=\{0,1\}$), monotonicity can be stated succinctly as
$$ A^1 \geq A^0 \text{ with probability one}. $$
Monotonicity rules out the possibility that there are troublesome units in the population with $A^0=1$ but $A^1=0$. Such units are called `defiers' since they take treatment $A=1$ when not encouraged by the instrument (i.e., when $Z=0$), but take control $A=0$ when they are in fact encouraged (i.e., when $Z=1$). Thus monotonicity ensures that the population only comprises never-takers ($A^0=A^1=0$), always-takers ($A^0=A^1=1$), and compliers ($A^0=0, A^1=1$). Monotonicity can often be a reasonable assumption in practice, but not always; it has been discussed extensively in previous work, particularly for binary instruments (see \textcite{imbens2014instrumental} and discussion for a nice overview).

A natural way to extend monotonicity to the continuous instrument setting is as follows.

\medskip
\begin{assumption}[Monotonicity] \label{ass:4mono}
If $z' > z$ then $A^{z'} \geq A^z$ with probability one.
\end{assumption}
\medskip

Under Assumption \ref{ass:4mono}, no unit would ever change from treatment to control with an increase in the instrument value; increasing the instrument can either encourage treatment over control or have no effect at all, but it cannot discourage treatment relative to lesser instrument values. Thus the population still comprises never-takers, always-takers, and compliers, but with continuous instruments the compliers can be further partitioned into compliers at given instrument values. In particular, a complier at $Z=z$ would be a unit for which $A^z=1$ but $A^{z-\delta}=0$ for any $\delta>0$. {We use a population version of monotonicity for simplicity, but only a covariate-specific conditional version is necessary when targeting conditional treatment effects \autocite{tan2006regression}.} The above continuous version of monotonicity has been employed and discussed  by \textcite{glickman2000derivation} and  \textcite{vytlacil2002independence}, for example. Importantly, these authors showed that (when coupled with standard identifying assumptions to be discussed shortly) the above monotonicity assumption is equivalent to the following latent threshold model. 

\medskip
\begin{assumptionprime}[Latent Threshold] \label{ass:4thresh}
$A^z = \one(z \geq T)$ for an unobserved random threshold $T$.
\end{assumptionprime}
\medskip

Under the latent threshold model, each complier has some instrument value at which they are encouraged to take treatment, while for any lesser value they would take control. Larger values of the threshold $T$ indicate units that are less willing to take treatment, i.e., less susceptible to encouragement by the instrument. We can thus define the latent threshold $T$ as 
$$ T = \begin{cases} -\infty & \text{if $A^z=1$ for all $z$ (always-takers)}  \\ \inf \{ z  : A^z =1 \} & \text{if $A^{z'}>A^z$ for some $z'>z$ (compliers)} \\ \infty & \text{if $A^z=0$ for all $z$ (never-takers).} \end{cases}$$
We refer to \textcite{vytlacil2002independence} for further discussion and detail.

\subsection{Estimand \& Identification}

In this paper our goal is estimation and inference for the local instrumental variable (LIV) curve, which we define as
\begin{equation} \label{eq:liv}
\gamma(t,\bv) = \E(Y^1 - Y^0 \mid T=t, \bV=\bv) .
\end{equation}
This is the average treatment effect among those with latent threshold $T=t$ (and value $\bV=\bv$ for some arbitrary baseline covariate subset), i.e., the effect among those units with $\bV=\bv$ who would be encouraged to take treatment right when the instrument passes $Z=t$ but not for lesser values. A fully conditional version of the LIV curve with $\bV=\bX$ was proposed in the latent index or selection model framework by \textcite{heckman1997instrumental}, and discussed in detail by \textcite{heckman1999local} and \textcite{heckman2005structural}. { Note that the variables $\bV$ are arbitrary, and no causal assumptions need to be made about them; this is because we are estimating effect modification (how effects vary with $\bV$) and not mediation (how effects would change if we set $\bV$ to certain values). }

{We study the LIV curve for continuous IVs since such IVs are very common in practice and since novel theory is required; an analogous LIV effect  $\E(Y^1 - Y^0 \mid A^t > A^{t-1}, \bV)$  for discrete multivalued instruments could be studied by adapting results from \textcite{tan2006regression, ogburn2015doubly}. However, for continuous instruments, the LIV effect curve is substantially different from the local average treatment effect (LATE) studied by \textcite{angrist1996identification}, \textcite{tan2006regression}, \textcite{ogburn2015doubly} and others. Specifically, the continuous IV extension of the LATE is the three-dimensional effect surface $\E(Y^1-Y^0 \mid A^z>A^{z'}, \bV=\bv)$ for values $z>z'$, which represents the effect among those who would take treatment at $Z=z$ but not $Z=z'$. In contrast, for continuous instruments, the LIV parameter in \eqref{eq:liv} is an arguably more interpretable (and easier to model) curve representing the effect among those ``at the margin'' who would take treatment at $Z=t$ but not $Z<t$. These and other differences between the LATE and LIV are discussed in more detail by \textcite{heckman1999local, heckman2005structural}. Further, even more substantial differences arise after identification, as discussed below. }

\begin{remark}
In Heckman's selection model framework, the LIV curve in \eqref{eq:liv} with $\bV=\bX$ was termed the ``marginal treatment effect'', and its observed data counterpart the ``local IVs'' estimand (after employing identifying assumptions). We use the latter in both cases, since ``marginal'' is often used to mean ``averaged'' instead of ``at the margin''.
\end{remark}

For identification we rely on standard IV assumptions, which have been employed for example by \textcite{angrist1996identification}, \textcite{tan2006regression}, \textcite{ogburn2015doubly}, and others; useful overviews and discussions are given by \textcite{hernan2006instruments}, \textcite{imbens2014instrumental} (with discussion), and \textcite{baiocchi2014instrumental}.

\medskip
\begin{assumption}[Consistency] \label{ass:4consistency}
$A=A^Z$ and $Y=Y^A$ with probability one.
\end{assumption}
\begin{assumption}[Positivity] \label{ass:4positivity}
$(z,\bx) \in \supp(Z,\bX)$ if $\bx \in \supp(\bX)$.
\end{assumption}
\begin{assumption}[Unconfoundedness of $Z$] \label{ass:4unconfound}
$(Y^z,A^z) \ind Z \mid \bX$.
\end{assumption}
\begin{assumption}[Exclusion Restriction] \label{ass:4exclusion}
$Y^{za}=Y^a$ with probability one.
\end{assumption}
\medskip

Consistency means potential treatments $A^z$ and outcomes $Y^a$ are uniquely defined by a unit's own instrument and treatment levels, respectively, and not by others' levels (i.e., no interference), and also not by the way the instrument or treatment are administered (i.e., no different versions). Positivity says that the instrument is not deterministic, in the sense that every unit has some chance of receiving each level of the instrument, regardless of covariates. Unconfoundedness says that the instrument is essentially randomized once we condition on covariates, i.e., that it is unrelated to potential outcomes $Y^z=Y^{zA^z}$ and treatments $A^z$ under different instrument values $Z=z$. The exclusion restriction says that the instrument only affects outcomes through treatment. Assumptions \ref{ass:4consistency}--\ref{ass:4exclusion} can hold by design in trials where the instrument is externally randomized by investigators, but in observational studies these assumptions are typically untestable and require justification based on subject matter.

Finally we also employ the following regularity conditions on the latent threshold distribution and LIV curve.

\medskip
\begin{assumption}[Instrumentation] \label{ass:4instrument}
$\inf_{t} p(t \mid \bv) > 0$.
\end{assumption}
\begin{assumption}[Continuity] \label{ass:4continuity}
$T$ is continuously distributed and $\gamma(t,\bv)$ is continuous in $t$.
\end{assumption}
\medskip

Instrumentation means there are at least some units whose treatment would be affected by the value of the instrument (i.e., some who would take treatment when the instrument passes $Z=t$); for now we leave the set over which the infimum is taken ambiguous. 
{
\begin{remark}
A caveat is in order regarding Assumptions 3 and 6: positivity and instrumentation are particularly strong requirements for continuous instruments. Positivity may be violated, for example, if some units simply have no chance at receiving IV levels far away from what they actually received; e.g., in the analysis in Section 5, it may be implausible that some patients who actually live very close to high-level ICUs could ever live very far from them. Similarly, instrumentation can be violated with weak instruments, since moving the IV past certain levels simply may not affect any units' treatment status. Positivity violations can be partially ameliorated by an appropriate choice of weight function (as discussed in Section 3.1), and instrumentation violations by an appropriate choice of the set $\mathcal{T}$ on which the LIV curve is to be estimated; however neither is a panacea. See \textcite{petersen2010diagnosing, westreich2010invited, stock2012survey, baiocchi2014instrumental} and others for relevant discussion of these issues in related settings. Extensions dealing with violations of these assumptions are not considered here. 
\end{remark} }

The following theorem indicates that the LIV curve can be identified with observed data, under the above assumptions.

\begin{theorem}
\label{thm:liv_id}
Suppose Assumption \ref{ass:4thresh} holds. Let $\mathcal{T} \subset \supp(Z)$ denote a compact set on which we wish to identify $\gamma(t,\bv)$. If Assumptions \ref{ass:4consistency}--\ref{ass:4exclusion} hold for all $z \in \mathcal{T}$ and Assumptions \ref{ass:4instrument}--\ref{ass:4continuity} hold for all $t \in \mathcal{T}$, then the LIV curve is identified for any $t \in \mathcal{T}$ by
\begin{equation}
\label{eq:liv_id}
\gamma(t,\bv) =\frac{ \frac{\partial}{\partial z} \E\{ \E(Y \mid \bX,Z=z) \mid \bV=\bv\}}{ \frac{\partial}{\partial z} \E\{ \E(A \mid \bX,Z=z) \mid \bV=\bv\} } \bigg|_{z=t}.
\end{equation}
\end{theorem}

A proof of Theorem \ref{thm:liv_id} is given in the Supplementary Materials; the logic follows as in more standard settings where $Z$ is discrete and $\bV=\bX$. Importantly, the LIV curve can only be identified on subsets of $\supp(Z)$; thus as in the binary instrument setting, we cannot identify effects for never-takers or always-takers with $T=\pm\infty$. {As discussed in more detail in the next section, the ratio-of-derivatives structure of the LIV curve makes its study particularly interesting from a theoretical perspective, especially relative to the LATE parameter studied by \textcite{ogburn2015doubly} and others.}

\begin{remark}
From this point forward,  $\gamma(t,\bv)$ will denote the observed data expression in \eqref{eq:liv_id}, which represents the causal effect given in \eqref{eq:liv} under Assumptions \ref{ass:4consistency}--\ref{ass:4continuity} as described in Theorem \ref{thm:liv_id}. Of course, if the conditions of Theorem \ref{thm:liv_id} do not hold then \eqref{eq:liv_id} may represent something other than the aforementioned causal effect. For example, if only Assumptions 2--4 hold, then we can only think of the instrument as an unconfounded continuous exposure (or dose), and $\gamma(t,\bv)$ would represent the ratio of derivatives of the dose-response curves $\E(Y^z \mid \bV=\bv)$ and $\E(A^z \mid \bV=\bv)$.
\end{remark}

\section{Main Results}

In this section we develop semiparametric theory for models of the LIV curve defined in \eqref{eq:liv_id}, use this theory to develop novel estimators (including inverse-probability-weighted, regression, and doubly robust estimators), describe asymptotic properties, and finally present cross-validation methods for model selection in high-dimensional settings.

\subsection{Semiparametric Theory}

Suppose we have a parametric model for the LIV curve, which we write as $\gamma(t,\bv;\bpsi)$ for some finite-dimensional $\bpsi \in \R^q$. Importantly, we do not assume this model is necessarily correct, and instead follow \textcite{neugebauer2007nonparametric}, \textcite{rosenblum2010targeted}, and others in using a working model approach, by formulating our estimand as the projection of the true curve $\gamma(t,\bv)$ onto the posed working model. Specifically, we use the weighted least squares projection given by
\begin{equation}
\bpsi_0 = \argmin_{\bpsi \in \R^q} \ \E\Big[ w(T,\bV) \{ \gamma(T,\bV) - \gamma(T,\bV;\bpsi) \}^2 \Big] ,
\label{eq:projection}
\end{equation}
where $w(t,\bv)$ is some user-specified weight function. {We use the $L_2$ loss-based projection for its convenience and familiarity, but it would be worthwhile to develop results for other loss functions in future work.} 

The projection approach warrants some discussion. Whether to use a model like $\gamma(t,\bv;\bpsi)$ only for projections, or to assume it is actually correct, can be viewed as a bias-variance trade-off. If the model happens to be correct, then both the projection and model-based approaches will yield valid estimates of the true function, though projection estimators will not generally be fully efficient (depending on the choice of weight function). However, if the posited model is incorrect, then the model-based approach is technically no longer valid and can be difficult to interpret, since it may not correspond to an interpretable projection (e.g., least-squares); on the other hand, the projection approach is still well-defined and represents a best-fitting summary measure. 

The projection approach essentially formalizes how models are often viewed as approximations in practice. However, we note that there are some distinctions between using approximations in causal versus predictive settings. In causal settings one often cares about understanding mechanisms or making beneficial treatment recommendations; when the model is incorrect, different projections can vary in their usefulness in these respects. And the costs of an inaccurate projection may be quite different for these causal goals (either smaller or larger), relative to usual predictive goals of minimizing prediction error. 

{Another important issue in using projections is choosing the weight $w(t,\bv)$. Ideally there will be some subject matter justification for learning about specific parts of the effect curve. If not, in theory one could use a uniform weight that assigns mass equally across the support, but in practice this could lead to poor efficiency. We have found that weights based on the instrument density, e.g., $w(t,\bv)=p(Z=t, \bv)$ or $w(t,\bv)=w(t)=p(Z=t)$, work well when no particular weight function is preferred based on substantive concerns. Of course if $\gamma$ is correctly specified, all weights yield consistent estimators that only vary in efficiency.}

Note that the projection parameter in \eqref{eq:projection} depends on the distribution of the latent threshold $T$. Although this threshold is not observed directly, its distribution is identified in the observed data (under Assumptions \ref{ass:4mono}--\ref{ass:4continuity}). For example when $t \in \supp(Z)$ we have
\begin{equation}
p(t \mid \bv) =  \frac{\partial}{\partial z} \E\{\E(A \mid \bX, Z=z) \mid \bV=\bv\} \Bigm|_{z=t}
\end{equation}
Importantly, the above expression for the threshold density equals the denominator of  $\gamma(t,\bv)$ given in Theorem \ref{thm:liv_id}. 

After characterizing the parameter of interest in terms of observed data as in \eqref{eq:projection}, based on the expression in \eqref{eq:liv_id}, it is possible to estimate it using any number of approaches, such as parametric or nonparametric maximum likelihood, or Bayesian methods. In our setting, however, semiparametric approaches have a number of important advantages. First, they can incorporate information about the instrument mechanism, which may be better understood or easier to model than the outcome and treatment mechanisms (which is what a likelihood-based approach would rely on modeling). Second, they allow for double robustness, which means consistent estimation of $\bpsi$ is possible as long as either the instrument mechanism or the treatment/outcome mechanisms are correctly modeled (not necessarily all three, so either the instrument or the treatment and outcome models can be misspecified). And third, semiparametric doubly robust approaches allow for fast root-n rates of convergence for the parameter of interest $\bpsi$, even when nuisance functions are estimated at slower rates, e.g., using flexible data-adaptive or machine learning methods. Thus these estimators are less sensitive to the curse of dimensionality; this phenomenon was noted recently for example by \textcite{van2014higher} and \textcite{chernozhukov2016double}, who refer to it as orthogonality. Nonetheless our results also lead to novel inverse-weighting and regression-based estimators; the latter in particular might be preferable in small samples.

A crucial aspect of developing semiparametric theory and corresponding estimators for a given problem involves characterizing the possible influence functions, and in particular finding the efficient influence function. Many details on semiparametric theory are available elsewhere \autocite{bickel1993efficient, van2003unified, tsiatis2006semiparametric, kennedy2016semiparametric}, so we give only a brief review here. Any regular asymptotically linear estimator minus its target parameter can be expressed as the empirical average of its so-called influence function plus an $o_\Pb(1/\sqrt{n})$ error term. Viewed as elements of a Hilbert space of mean-zero finite-variance functions equipped with covariance norm, the influence functions under a given model lie in the orthogonal complement of the nuisance tangent space. The efficient influence function can then be defined as the influence function with smallest variance, the projection of any influence function onto the tangent space of scores, or as a particular pathwise derivative. The efficient influence function is especially important in practice because its variance is the semiparametric efficiency bound (thus providing a benchmark for efficient estimation), and because it can be used to construct estimators that are doubly robust and potentially semiparametric efficient.

A major challenge in deriving semiparametric theory for the projection parameter in \eqref{eq:projection} is its complexity; namely, it is a weighted projection of a ratio of derivatives of regression functions that are partially marginalized. {We conjecture that this may be why a doubly robust estimator for the LIV curve has yet to appear in the literature. In general, such complex structure would yield a complicated efficient influence function involving derivatives of regression functions, making corresponding estimators very difficult to compute. A major contribution of our work is an expression for the efficient influence function that only involves derivatives based on the known (and generally more simple) model and weight functions, rather than unknown complex and high-dimensional nuisance functions. Although still complex, our formulation yields efficient estimators that are easier to construct in practice.} The next theorem gives the efficient influence function for the parameters of the LIV curve projection.

\begin{theorem} \label{thm:4eif}
Suppose the weight function $w(t,\bv)$ is continuously differentiable in $t$ and satisfies $w(t,\bv)=0$ for $t \notin \text{int}(\mathcal{T})$, with the set $\mathcal{T} \subset \supp(Z)$ defined as in Theorem \ref{thm:liv_id}. Also assume that partial derivatives (with respect to $\bpsi$ and $t$) of the working model $\gamma(t,\bv;\bpsi)$ exist and are continuous. Then, under a nonparametric model, the efficient influence function for $\bpsi$ defined in  \eqref{eq:projection} is proportional to
\begin{align} \label{eq:eif}
\boldsymbol\varphi(\bO;\bpsi,\boldsymbol\eta) &=  \int_\mathcal{T} \Big\{ \bg_1(t,\bV;\bpsi) \E(A \mid \bX,Z=t) - \bg_2(t,\bV;\bpsi) \E(Y \mid \bX,Z=t) \Big\} \ dt  \\
& \hspace{.4in} + \bg_1(Z,\bV; \bpsi) \left\{ \frac{A-\E(A \mid \bX,Z)}{p(Z \mid \bX)} \right\} - \bg_2(Z,\bV;\bpsi) \left\{ \frac{Y - \E(Y \mid \bX,Z)}{p(Z \mid \bX)} \right\} \nonumber  
\end{align}
where $\boldsymbol\eta=(\pi,\lambda,\mu)$ denotes the nuisance functions defined in Section \ref{4data}, and $\bg_1$ and $\bg_2$ are the $(q \times 1)$ vectors
\begin{align*}
\bg_1(z,\bv;\bpsi) &= \frac{\partial}{\partial t} \left\{ \frac{\partial}{\partial \bpsi^*} \gamma(t,\bv; \bpsi^*) \Big|_{\bpsi^*=\bpsi} w(t,\bv) \gamma(t,\bv;\bpsi) \right\}\Big|_{t=z} \\
\bg_2(z,\bv;\bpsi) &= \frac{\partial}{\partial t} \left\{ \frac{\partial}{\partial \bpsi^*} \gamma(t,\bv; \bpsi^*)  \Big|_{\bpsi^*=\bpsi} w(t,\bv)  \right\} \Big|_{t=z} .
\end{align*}
\end{theorem}

A proof of Theorem \ref{thm:4eif} is given in the Supplementary Materials. The first step is to define $\bpsi$ as the zero of a moment condition coming from the derivative of \eqref{eq:projection}. Then, to deal with the crucial difficulty that this moment condition involves complex derivatives of partially marginalized treatment/outcome regression functions as in \eqref{eq:liv_id}, we use integration by parts to instead transfer the derivatives  to the known model $\gamma(t,\bv;\bpsi)$ and weight function $w(t,\bv)$. This allows us to write the moment condition as a somewhat more standard causal quantity, essentially equaling sums of the mean treatment/outcome under a stochastic intervention on the instrument, where the conditional instrument density is replaced with a uniform distribution on $\mathcal{T}$.  At this point one can proceed by showing that the function $\boldsymbol\varphi$ is the canonical gradient of the pathwise derivative of $\bpsi$. 

This explains why the influence function in Theorem \ref{thm:4eif} takes the form it does: it resembles that of a (uniform) stochastic intervention effect, as in for example \textcite{diaz2012population}, with the non-standard form of the functions $\bg_1$ and $\bg_2$ coming from the integration by parts trick.  The influence function as a whole can be viewed as the ``derivative term'' in a generalized von Mises expansion of the moment condition, as in for example \textcite{van2014higher}; alternatively, as in other causal inference and missing data problems, it can be viewed as consisting of an inverse-probability-weighted term (the added terms in the second line of \eqref{eq:eif}) plus an augmentation term (the first line and subtracted terms in the second line of \eqref{eq:eif}).

Importantly, using integration by parts to transfer derivatives from $P$-dependent quantities also means the influence function can be evaluated without analytical differentiation of the regression functions, which makes it much more practical for constructing and implementing estimators, as discussed in more detail in the next subsection. The user-specified weight function is required to vanish outside the interior of the set $\mathcal{T} \subset \supp(Z)$ since the LIV curve $\gamma(t,\bv)$ is not identified outside of $\mathcal{T}$ as discussed in Theorem \ref{thm:liv_id}. 

\subsection{Proposed Method}

Once we have derived the efficient influence function, we can use it to construct estimators that have numerous advantageous properties. A standard approach is to solve an estimating equation based on an estimated version of the efficient influence function; specifically we can use $\boldsymbol\varphi$ as an estimating function, with unknown nuisance functions replaced with estimates. 

Thus our proposed estimator for a given working model $\gamma(t,\bv;\bpsi)$ is given by $\boldsymbol{\hat\psi}$, defined as the solution in $\bpsi$ to the estimating equation
\begin{equation} \label{eq:4est}
 \Pn\{ \boldsymbol\varphi(\bO; \bpsi, \boldsymbol{\hat\eta}) \} = \mathbf{0} , 
\end{equation}
where $\boldsymbol{\hat\eta}=(\hat\pi,\hat\lambda,\hat\mu)$ are estimated versions of the three nuisance functions $\boldsymbol{\eta}=(\pi,\lambda,\mu)$. An inverse-probability-weighted estimator can be obtained by using $\hat\lambda=\hat\mu=0$, and a regression estimator can be obtained by using $\hat\pi=\infty$. Another option for constructing estimators based on influence functions is targeted minimum loss-based methodology \autocite{van2006targeted}, which yields plug-in estimators that respect the bounds of the parameter space. In our setting, our proposed estimating equation approach will also respect any such bounds, as long as the chosen working model does; it is also relatively straightforward to implement. 
{
\begin{remark}
For notational simplicity, the estimator proposed above uses nuisance estimates constructed from the entire sample (with asymptotic results in Section 3.3 relying on empirical process conditions). In Section 4 of the Supplementary Materials we present sample-splitting estimation, in the same spirit as \textcite{robins2008higher}, \textcite{zheng2010asymptotic},  \textcite{robins2013new}, \textcite{chernozhukov2016double}, and others, which does not require empirical process conditions and allows arbitrarily complex and adaptive nuisance estimators, e.g., random forests.
\end{remark} }

First consider the simple case where $\bV=\emptyset$ (i.e., effect modification is not of interest), and the LIV curve $\gamma(t)$ is projected onto a constant $\gamma(t;\bpsi)=\psi$. Here the target estimand $\psi$ is a simple weighted average of $\gamma(t)$, of the form
$$ \psi = \int_\mathcal{T} w^*(t) \gamma(t) \ dt $$
with weight $w^*(t)= w(t) p(t) / \int_\mathcal{T} w(t) p(t) \ dt$, {where $w(t)=w(t,\bv)$ since $\bV$ is empty}. Note that in general the parameter $\bpsi$ indexes a model for the threshold-dependent LIV curve $\gamma(t,\bv)$, but here it is a weighted average over $T$ because we have used a simple projection and are agnostic about whether the constant effect model is correct. In this case the quantity $\psi$ can also be viewed as the mean treatment effect (among compliers) in a population where the density of the latent threshold $T$ among compliers equals $w^*(t)$. Solving \eqref{eq:4est} leads to the ratio estimator
\begin{equation} \label{eq:wavg}
 \hat\psi = \frac{ \int_\mathcal{T} w'(t) \hat{m}(t) \ dt + \Pn \left\{ w'(Z) \frac{Y-\hat\mu(\bX,Z)}{\hat\pi(Z \mid \bX)} \right\}  } { \int_\mathcal{T} w'(t) \hat\ell(t) \ dt + \Pn \left\{ w'(Z)  \frac{A-\hat\lambda(\bX,Z)}{\hat\pi(Z \mid \bX)} \right\} } 
\end{equation}
where $\hat{m}(t)=\Pn\{\hat\mu(\bX,t) \}$ and $\hat\ell(t)=\Pn \{ \hat\lambda(\bX,t) \}$ are estimates of the marginalized regression functions from Section \ref{4data}. Thus $\hat\psi$ is an adjusted version of the regression-based plug-in estimator $\int_\mathcal{T} w'(t) \hat{m}(t) \ dt /  \int_\mathcal{T} w'(t) \hat\ell(t) \ dt$, where adding inverse-probability-weighted terms to the numerator and denominator is the adjustment required to obtain double robustness. 

More standard IV estimators are often computed with a two-stage least squares approach, where in the first stage the treatment variable is regressed on the instrument (and covariates) and then in the second stage the outcome is regressed on the predicted values from the first stage (and covariates). In fact, the weighted average estimator in \eqref{eq:wavg} can also be constructed with a modified version of such a two-stage least squares approach; this may make it more amenable to practical use. Specifically, the following modified two-stage least squares procedure can be used to compute the weighted average estimator  (using pseudo- instrument, treatment, and outcome $w'(Z)$, $A^*$, and $Y^*$ respectively):

\begin{enumerate}
\item Regress $A^* = \frac{A - \hat\lambda(\bX,Z)}{\pi(Z \mid \bX)} + \frac{ \one\{w'(Z) \neq 0\}}{w'(Z)} \int_\mathcal{T} w'(t) \hat\lambda(\bX,t) \ dt$ on $w'(Z)$ without an intercept, and obtain predicted values $\hat{A}^*$. 
\item Regress $Y^* = \frac{Y - \hat\mu(\bX,Z)}{\pi(Z \mid \bX)} + \frac{ \one\{w'(Z) \neq 0\}}{w'(Z)} \int_\mathcal{T} w'(t) \hat\mu(\bX,t) \ dt$ on $\hat{A}^*$, without an intercept.
\end{enumerate}
Then the coefficient in front of $\hat{A}^*$ in the second stage equals $\hat\psi$ from \eqref{eq:wavg}.

Closed-form estimators are also available even when effect modification is of interest, as long as we project onto linear models of the form $\gamma(t,\bv;\bpsi)=\bh(t,\bv)^\T \bpsi$, for some known mapping $\bh: \mathcal{T} \times \supp(\bV) \rightarrow \R^q$. Specifically, in such cases the estimator $\boldsymbol{\hat\psi}$ defined as the solution to \eqref{eq:4est} is given by
\begin{align*}
\boldsymbol{\hat\psi} &= \Pn \left[ \bg_1^*(Z,\bV) \left\{ \frac{A-\hat\lambda(\bX,Z)}{\hat\pi(Z \mid \bX)} \right\} + \int_\mathcal{T} \bg_1^*(t,\bV) \hat\lambda(\bX,t) \ dt \right]^{-1} \\
& \hspace{.4in} \times \Pn \left[ \bg_2(Z,\bV) \left\{ \frac{Y-\hat\mu(\bX,Z)}{\hat\pi(Z \mid \bX)} \right\} + \int_\mathcal{T} \bg_2(t,\bV) \hat\mu(\bX,t) \ dt \right]
\end{align*}
where $\bg_1^*(z,\bv) = \frac{\partial}{\partial t} \{ \bh(t,\bv) w(t,\bv) \bh(t,\bv)^\T\}|_{t=z}$, and $\bg_2(z,\bv)=\frac{\partial}{\partial t} \{ \bh(t,\bv) w(t,\bv)\}|_{t=z}$ is as defined in Theorem \ref{thm:4eif}. Closed-form expressions will typically not be available for estimators in general non-linear models; however, since such estimators are still defined as estimating equation-based Z-estimators, they can be computed with standard software (for example, one could use the \verb+optim+ function in R). Variance estimation and confidence interval construction will be discussed in the next section.

\subsection{Asymptotic Theory}

In this section we discuss the asymptotic properties of our proposed estimation approach. In particular we show that our estimator is doubly robust, and that if the nuisance functions are estimated well enough it is asymptotically normal and efficient. Further, asymptotic normality and efficiency are possible even after flexible machine learning-based covariate adjustment. (Our results equally apply to estimators that only solve the efficient influence function estimating equation asymptotically, up to order $o_\Pb(1/\sqrt{n})$, such as targeted minimum loss-based estimators.) {An analogous result holds for a sample-splitting version of the proposed estimator, as detailed in Section 6 of the Supplementary Materials.}

\begin{theorem} \label{thm:4asymptotics}
Assume that:
\begin{enumerate}
\item $(\boldsymbol{\hat\psi},\boldsymbol{\hat\eta}) \inprob (\bpsi_0,\boldsymbol{\overline\eta})$, where $\boldsymbol{\overline\eta}=(\overline\pi,\overline\lambda,\overline\mu)$ with either $\overline\pi=\pi_0$ or $(\overline\lambda,\overline\mu)=(\lambda_0,\mu_0)$.
\item The sequence of functions $\boldsymbol{\hat\varphi}_n = \boldsymbol\varphi(\cdot;\boldsymbol{\hat\psi},\boldsymbol{\hat\eta})$ and its limit $\boldsymbol\varphi_0=\boldsymbol\varphi(\cdot;\boldsymbol{\psi_0},\boldsymbol{\overline\eta})$ are contained in a Donsker class with $||\boldsymbol{\hat\varphi}_n - \boldsymbol\varphi_0||_2=o_\Pb(1)$.
\item The map $\bpsi \rightarrow \E\{ \boldsymbol\varphi(\bO;\bpsi,\boldsymbol\eta) \}$ is differentiable at $\bpsi_0$ uniformly in $\boldsymbol\eta$ (around $\boldsymbol{\overline\eta}$), with invertible derivative matrix $\mathbf{D}(\bpsi_0,\boldsymbol\eta) \rightarrow \mathbf{D}(\bpsi_0,\boldsymbol{\overline\eta}) \equiv \mathbf{D}_0$.
\end{enumerate}
Then the proposed estimator is consistent with rate of convergence
$$ || \boldsymbol{\hat\psi} - \bpsi_0 || = O_\Pb\left\{ 1/\sqrt{n} + ||\hat\pi - \pi_0||_2 \Big( ||\hat\lambda - \lambda_0 ||_2 + ||\hat\mu - \mu_0 ||_2 \Big) \right\} . $$
Suppose further that:
\begin{enumerate}
\item[(d)] $||\hat\pi - \pi_0||_2 ( ||\hat\lambda - \lambda_0 ||_2 + ||\hat\mu - \mu_0 ||_2 )=o_\Pb(1/\sqrt{n})$.
\end{enumerate}
Then the proposed estimator is asymptotically normal with
$$ \sqrt{n}(\boldsymbol{\hat\psi} - \bpsi_0) \indist N\Big(\mathbf{0}, \E[\{ \mathbf{D}_0^{-1} \boldsymbol\varphi(\bO;\bpsi_0,\boldsymbol\eta_0) \}^{\otimes 2}] \Big) , $$
and thus semiparametric efficient.
\end{theorem}

A proof of Theorem \ref{thm:4asymptotics} is given in the Supplementary Materials; it follows from standard Z-estimator theory and empirical process results \autocite{van1996weak,  van2002semiparametric}. The first condition indicates the double robustness of our approach, since some of the nuisance estimators $\boldsymbol{\hat\eta}=(\hat\pi,\hat\lambda,\hat\mu)$ can be misspecified. Specifically, as long as either $\hat\pi$ or $(\hat\lambda,\hat\mu)$ is consistent, then the estimator $\boldsymbol{\hat\psi}$ will be as well. This gives analysts two chances at consistency, and is particularly important in the IV setting since it can be easier to model the instrument density $\pi$ than the two regression functions $(\lambda,\mu)$. 

Conditions (b)--(c) of Theorem \ref{thm:4asymptotics} are standard regularity conditions for M- and Z-estimators  \autocite{van1996weak, van2000asymptotic, van2002semiparametric}. {Condition (b) restricts the flexibility of the nuisance estimators (and their limits), but Donsker classes still cover many complex functions.} For example, parametric Lipschitz functions are Donsker, but so are many more complicated function types such as infinite-dimensional smooth functions with bounded partial derivatives, VC classes, Sobolev classes, and functions with bounded uniform sectional variation, as well as convex combinations and Lipschitz transformations of any these classes. More discussion and examples can be found in Sections 2.6--2.7 of \textcite{van1996weak} and Examples 19.6--19.12 of \textcite{van2000asymptotic}, as well as in \textcite{kennedy2016semiparametric}. Condition (b) is important because it means we do not have to rely on restrictive parametric models to estimate the potentially complicated and high-dimensional nuisance functions $(\pi,\lambda,\mu)$, and can instead use more flexible data-adaptive methods. Condition (b) can also be weakened in various ways. For example, the Donsker condition really only needs to hold in a shrinking neighborhood of $(\bpsi_0,\boldsymbol{\overline\eta})$, or with high probability as $n \rightarrow \infty$; alternatively we could formulate Condition (b) in terms of weaker entropy or bracketing conditions. {Importantly, in Section 6 of the Supplementary Materials, we discuss how the sample-splitting estimator mentioned earlier (and defined in Section 4 of the Supplementary Materials) can do away with empirical process conditions entirely, allowing arbitrarily flexible nuisance estimators such as random forests}. The differentiability in Condition (c) is standard and required to use a delta method-type result (note that the influence function need not be differentiable itself, only its expectation).

Under Conditions (a)--(c) of Theorem \ref{thm:4asymptotics}, the proposed estimator is consistent with rate of convergence given by $1/\sqrt{n} + ||\hat\pi - \pi_0||_2 ( ||\hat\lambda - \lambda_0 ||_2 + ||\hat\mu - \mu_0 ||_2 )$. Again the double robustness is apparent since consistency (i.e., $|| \boldsymbol{\hat\psi} - \bpsi_0 || = o_\Pb(1)$) follows as long as either the instrument density is consistently estimated, i.e., $||\hat\pi - \pi_0||_2 = o_\Pb(1)$, or the treatment/outcome regressions are, i.e., $(||\hat\lambda - \lambda_0 ||_2 + ||\hat\mu - \mu_0 ||_2)=o_\Pb(1)$. 
Importantly, however, the result also shows how double robustness is useful even apart from giving two chances at consistency; in particular, if we estimate the regression functions $(\lambda,\mu)$ at slower rates, double robustness gives us a chance to obtain faster rates for $\boldsymbol{\hat\psi}$ by consistently estimating $\pi$, and vice versa. 

For example, if Condition (d) holds so that $||\hat\pi - \pi_0||_2 ( ||\hat\lambda - \lambda_0 ||_2 + ||\hat\mu - \mu_0 ||_2 )=o_\Pb(1/\sqrt{n})$, i.e., effects of nuisance estimation are asymptotically negligible, then the estimator $\boldsymbol{\hat\psi}$ is root-n consistent, asymptotically normal, and semiparametric efficient. Note Condition (d) can hold even if the nuisance functions are estimated at slower than parametric root-n rates, so that efficient estimation and valid inference is possible for $\bpsi$ even if we use machine learning-based covariate adjustment, via flexible estimation of the nuisance functions $(\pi,\lambda,\mu)$. For example, if the nuisance functions $(\pi,\lambda,\mu)$ are all estimated at faster than $n^{1/4}$ rates, so that $||\hat\pi - \pi_0||_2 = ||\hat\lambda - \lambda_0 ||_2 = ||\hat\mu - \mu_0 ||_2 =o_\Pb(n^{-1/4})$, then Condition (d) holds since $o_\Pb(n^{-1/4})o_\Pb(n^{-1/4})=o_\Pb(1/\sqrt{n})$. {Thus in this case the rate for $\hat{\bpsi}$ is faster than that of the nuisance estimators  (here, it is the square of the slower nuisance rates).} Such $n^{1/4}$ rates are possible in various flexible models; for instance, under some conditions \autocite{horowitz2009semiparametric} generalized additive model estimators can obtain rates of the form $O_\Pb(n^{-2/5})$, which is $o_\Pb(n^{-1/4})$ since 
$$R_n=O_\Pb(n^{-2/5}) \implies n^{1/4}R_n = n^{-3/20} n^{2/5} R_n=O_\Pb(n^{-3/20})=o_\Pb(1). $$ 
Condition (d) can also hold if one of $\pi$ or $(\lambda,\mu)$ is estimated with a correctly specified parametric model and the other is merely estimated consistently. 

If Condition (d) holds, confidence intervals can be constructed with the bootstrap, or using a direct estimate of the asymptotic variance given in Theorem \ref{thm:4asymptotics}, such as
$$  \Pn[ \{ \mathbf{\hat{D}}^{-1}\boldsymbol\varphi(\bO;\boldsymbol{\hat\psi},\boldsymbol{\hat\eta})\}^{\otimes 2}] $$
where $\mathbf{\hat{D}}= \Pn\{ \partial \boldsymbol\varphi(\bO;\boldsymbol{\psi},\boldsymbol{\hat\eta}) / \partial \bpsi^\T\}|_{\bpsi=\boldsymbol{\hat\psi}}$ is an estimate of the derivative matrix from Condition (c) of Theorem \ref{thm:4asymptotics}. {For completeness}, we note that if parametric models are used to estimate all three nuisance functions (and either $\pi$ or $(\lambda,\mu)$ are correctly modeled, but not both) then the bootstrap would still be valid even though Condition (d) fails, since then the contribution from nuisance estimation is asymptotically linear (an analytic expression could also be derived, since $\boldsymbol{\hat\psi}$ and the estimated nuisance parameters solve a large system of estimating equations). {However, as we note in the Introduction, parametric nuisance models are likely to be misspecified, except for $\pi$ if $Z$ is externally randomized.} Inference is somewhat more delicate in a truly doubly robust but nonparametric setting, where one nuisance estimator can be misspecified but $(\hat\pi,\hat\lambda,\hat\mu)$ are estimated flexibly; we leave this to future work. 

\subsection{Model Selection}

To this point we have presumed that we have an a priori model $\gamma(t,\bv;\bpsi)$, which either represents the truth or a low-dimensional projection. When such a priori models are not available, we might instead aim to learn the form of $\gamma(t,\bv)$ from data. Thus in this section we propose a doubly robust cross-validation approach for model selection. Model selection is an important issue in causal inference in general, but this is especially the case for the LIV curve, since the latent threshold $T$ is continuous; thus saturated parametric models are not possible, even when effect modification is not of interest (i.e., $\bV=\emptyset$). In this section we derive the efficient influence function for the risk of a given candidate estimator, and show how it can be used as a doubly robust loss function in the cross-validation framework developed by \textcite{van2003cross}. {In particular, our cross-validation approach could be used to select among working models whose complexity increases with sample size, yielding doubly robust yet nonparametric estimators of the local IV curve, in the same spirit as \textcite{robins2001inference} and \textcite{kennedy2017nonparametric}.}

If we knew the true LIV curve and true distribution $P$, we could evaluate the performance of a given estimator $\hat\gamma_k$ by computing the mean squared error risk
$R^*(\hat\gamma_k) = \int_\mathcal{V} \int_\mathcal{T} w(t,\bv) \{ \gamma(t,\bv) - \hat\gamma_k(t,\bv) \}^2 \ dP(t,\bv)$. 
Alternatively, if we only wanted to compare or rank a set of candidate estimators $\{ \hat\gamma_k : k \in \mathcal{K}\}$, we could use the pseudo-risk
\begin{equation} \label{eq:4risk}
R(\hat\gamma_k) = \int_\mathcal{V} \int_\mathcal{T} w(t,\bv) \Big\{ \hat\gamma_k(t,\bv)^2 - 2\gamma(t,\bv) \hat\gamma_k(t,\bv)  \Big\} \ dP(t,\bv) ,
\end{equation}
since $R(\hat\gamma_k)=R^*(\hat\gamma_k) - \E\{w(T,\bV) \gamma(T,\bV)^2\}$ is simply a shifted version of the mean squared error $R^*(\hat\gamma_k)$, and the shift does not depend on the candidate estimator $\hat\gamma_k$. In standard cross-validation it is possible to estimate risk unbiasedly, without worrying about nuisance function estimation; in contrast, in our setting the risk parameter $R(\hat\gamma_k)$ depends on complex nuisance functions via the curve $\gamma(t,\bv)$ and the distribution of the threshold $T$. Thus estimation of the risk $R(\hat\gamma_k)$ itself requires nuisance estimation, and in fact we can treat $R(\hat\gamma_k)$ as a parameter in its own right, for which we can develop semiparametric theory and estimators. Thus in the next theorem we give the efficient influence function for the risk $R(\gamma_k)$  for a given fixed candidate $\gamma_k$, and go on to show how to use this efficient influence function as a doubly robust loss function for cross-validation-based model selection. 

\begin{theorem} \label{thm:4eif2}
Consider the same setting and assumptions as in Theorem \ref{thm:4eif}. Under a nonparametric model, the efficient influence function for the risk $R(\gamma_k)$ defined in \eqref{eq:4risk} for a fixed candidate $\gamma_k$ is given by $L(\bO;\gamma_k,\boldsymbol\eta) - R(\gamma_k)$, for
\begin{align} \label{eq:4loss}
L(\bO;\gamma_k,\boldsymbol\eta) &= \int_\mathcal{T} \Big\{ f_1(t,\bV;\gamma_k) \E(Y \mid \bX,Z=t) - f_2(t,\bV;\gamma_k) \E(A \mid \bX, Z=t) \Big\} \ dt  \\
& \hspace{.4in} + f_1(Z, \bV; \gamma_k) \left\{ \frac{Y - \E(Y \mid \bX,Z)}{p(Z \mid \bX)} \right\}  - f_2(Z,\bV;\gamma_k) \left\{ \frac{A-\E(A \mid \bX,Z)}{p(Z \mid \bX)} \right\}  \nonumber 
\end{align}
where $\boldsymbol\eta=(\pi,\lambda,\mu)$ are the nuisance functions from before, and $f_1$ and $f_2$ are defined as
\begin{align*}
f_1(z,\bv;\bpsi) &= 2\frac{\partial}{\partial t} \Big\{w(t,\bv) \gamma_k(t,\bv) \Big\}\Big|_{t=z} \\ 
f_2(z,\bv;\bpsi) &= \frac{\partial}{\partial t} \Big\{ w(t,\bv) \gamma_k(t,\bv)^2 \Big\} \Big|_{t=z} . 
\end{align*}
\end{theorem}

A proof of Theorem \ref{thm:4eif2} is given in the Supplementary Materials, and follows similar logic as the proof of Theorem \ref{thm:4eif}. We also show that $L(\bO;\gamma_k,\boldsymbol\eta)$ is a doubly robust loss function for the risk $R(\gamma_k)$ in the sense that $\E\{L(\bO;\gamma_k,\boldsymbol{\overline\eta})\}=R(\gamma_k)$ for nuisance function $\boldsymbol{\overline\eta}=(\overline\pi,\overline\lambda,\overline\mu)$ as long as either $\overline\pi=\pi_0$ or $(\overline\lambda,\overline\mu)=(\lambda_0,\mu_0)$, and not necessarily both. Thus we can use $L(\bO;\gamma_k,\boldsymbol\eta)$ as a doubly robust estimating function, similar to how we used $\boldsymbol\varphi(\bO;\bpsi,\boldsymbol\eta)$ in previous sections. However, since we typically do not have an independent sample to generate candidates $\hat\gamma_k$, we need to generate them from the same sample in which we estimate risk. Thus we can use sample-splitting  to prevent over-fitting.

In particular, we propose using the loss function in \eqref{eq:4loss} for doubly robust model selection following the general approach of \textcite{van2003cross}. This requires some new notation. Let $\bS=(S_1,...,S_n)$ denote a random variable independent of the sample that splits the data into training ($S_i=0$) and test ($S_i=1$) sets. For example standard $v$-fold cross-validation arises by allowing the split variable $\bS$ to take $v$ different values $\{ \bS_1,...,\bS_v\}$, each with equal probability $1/v$, where $\sum_i S_{iv}=n/v$ for all $v$ and $\sum_v S_{iv}=1$ for all $i$, so that test sets are all of size $n/v$ and each unit is only used in one test set. Further define $\Pb_\bs^0$ and $\Pb_\bs^1$ as the sub-empirical distributions for the training data $\{i: S_i=0\}$ and test data $\{i: S_i=1\}$, respectively, for a given split $\bS=\bs$. Therefore, for example, $\boldsymbol{\hat\eta}(\Pb_\bs^0)$ denotes the nuisance function estimates based only on the training set data, and $\hat\gamma_k(\Pb_\bs^0)$ denotes the LIV curve estimate based only on the training set data (which also depends on the nuisance function estimates constructed from the training data).

The cross-validation selection approach of \textcite{van2003cross} is very similar to standard cross-validation, but incorporates extra steps for nuisance function estimation; it proceeds as follows. For a given split $\bs$ and candidate estimator $\hat\gamma_k$, we first estimate the nuisance functions with the training data to obtain $\boldsymbol{\hat\eta}(\Pb_\bs^0)$, and then estimate the LIV curve with the training data to obtain $\hat\gamma_k(\Pb_\bs^0)$. Then the loss function $L$ can be evaluated for any observation $\bO_i$ based on these training estimates, and thus we do so on the test data $\Pb_\bs^1$ and compute the average, given by
$$ \hat{R}_\bs(\hat\gamma_k) = \int L\Big\{\bo; \hat\gamma_k(\Pb_\bs^0), \boldsymbol{\hat\eta}(\Pb_\bs^0) \Big\} \ d\Pb_\bs^1(\bo) , $$
which we call the estimated risk for candidate $k$ at the current split $\bs$. We repeat the above process for each split, average the split-specific risk estimates to get an overall risk estimate for candidate $k$, defined as $\hat{R}(\hat\gamma_k) = \E_\bS\{ \hat{R}_\bS(\hat\gamma_k) \}$, and finally repeat for each candidate $k \in \mathcal{K}$ and pick the one $\hat{k}$ that yields the smallest overall risk estimate $\hat{k}=\argmin_{k \in \mathcal{K}} \hat{R}(\hat\gamma_k)$. Hence the cross-validation selector can be written as
\begin{equation}
\hat{k} = \argmin_{k \in \mathcal{K}} \ \E_\bS \int L\Big\{\bo; \hat\gamma_k(\Pb_\bS^0), \boldsymbol{\hat\eta}(\Pb_\bS^0) \Big\} \ d\Pb_\bS^1(\bo) .
\end{equation}
 \textcite{van2003cross} gave conditions under which the risk $\hat{R}(\hat\gamma_{\hat{k}})$ of the above cross-validation selector is asymptotically equivalent to that of an oracle selector given by $\tilde{k} = \argmin_{k \in \mathcal{K}} \ \E_\bS \int L\{\bo; \hat\gamma_k(\Pb_\bS^0), \boldsymbol{\overline\eta} \} \ dP(\bo)$, 
along with corresponding finite-sample bounds. {One important condition is that the number of candidates  does not grow faster than polynomially with sample size, i.e., $|\mathcal{K}| \leq n^c$.} We refer to \textcite{van2003cross} for more details. {Note that semiparametric doubly robust post-selection inference is not considered here, but is an important avenue for future work}.

\section{Simulation Study}

To explore finite sample properties of our methods, we simulated from a model with
\begin{equation*}
\begin{gathered}
(Y^0, \bX )\sim N(\mathbf{0},\mathbf{I}_5), \\
Z \mid \bX, Y^0 \sim \text{TN}\{1.5 \times \text{sign}(\boldsymbol\alpha^T \bX),4,(-2,2)\}, \\ 
T \mid Z, \bX, Y^0 \sim  N(\boldsymbol\beta^T \bX+Y^0,1), \\
A = \one(Z \geq T), \
Y = Y^0 + A (\psi T),
\end{gathered}
\end{equation*}
where TN$\{\mu,\sigma^2,(l,u)\}$ denotes a truncated normal distribution with support $[l,u]$,  $\boldsymbol\alpha=(1,1,-1,-1)^\T$, $\boldsymbol\beta=(1,-1,-1,1)^\T$, and $\psi=1$. This setup satisfies the necessary identifying assumptions, and since $Y^1-Y^0 = \psi T$ it follows that the LIV curve is linear in the latent threshold $T$ (regardless of the conditioning set $\bV$). It can also be shown that the above setup implies
\begin{equation*}
\begin{gathered}
\E(A \mid \bX,Z) = \Phi\left(\frac{Z-\boldsymbol\beta^T \bX}{\sqrt{2}}\right), \\
\E(Y \mid \bX,Z) = \psi \left\{ (\boldsymbol\beta^T \bX) \Phi\left(\frac{Z-\boldsymbol\beta^T \bX}{\sqrt{2}}\right) - \sqrt{2} \phi\left(\frac{Z-\boldsymbol\beta^T \bX}{\sqrt{2}}\right) \right\},
\end{gathered}
\end{equation*}
so that the treatment regression $\lambda$ follows a probit model; the outcome regression $\mu$ is more complicated, but $\E(Y \mid \bX, Z,A)$ follows a particular generalized additive model. 

\begin{figure}[h!]
\begin{center}
\includegraphics[width=1\textwidth]{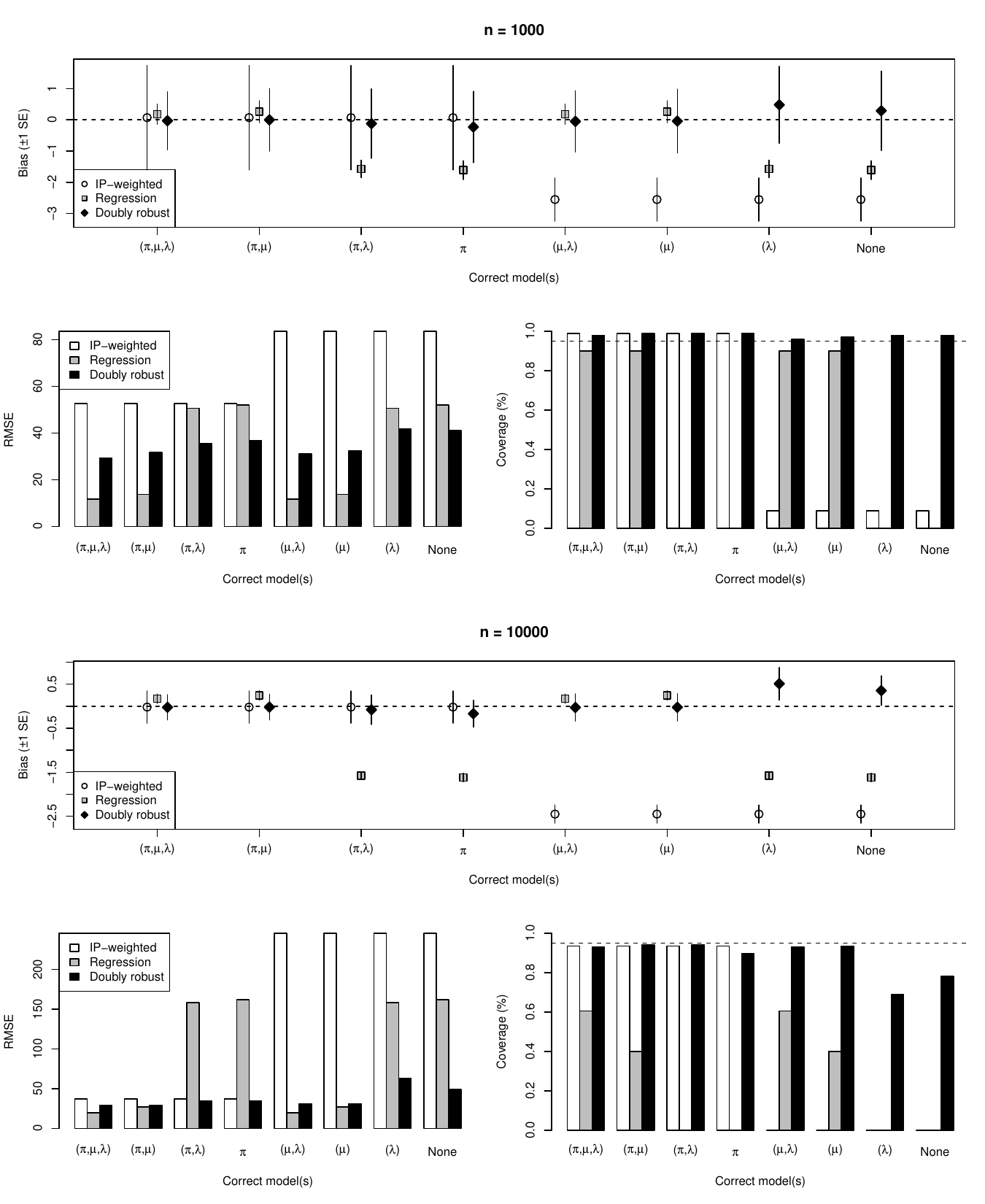}
\end{center}
\caption{Simulation results across 500 simulations: bias (with standard error), RMSE (scaled by $\sqrt{n}$), and bootstrap coverage (based on 100 bootstrap samples). \label{fig:simres}}
\end{figure}

We fit the proposed estimator, which in this case has a closed form given by
$$ \hat\psi = \frac{ \Pn \left[ \{ w(Z) + Zw'(Z) \} \left\{ \frac{Y-\hat\mu(\bX,Z)}{\hat\pi(Z \mid \bX)} \right\} \right]  + \int_\mathcal{T} \{w(t) + tw'(t)\} \hat{m}(t) \ dt}
{ \Pn \left[ \{ 2Z w(Z) + Z^2 w'(Z) \} \left\{ \frac{A-\hat\lambda(\bX,Z)}{\hat\pi(Z \mid \bX)} \right\} \right]  + \int_\mathcal{T} \{ 2t w(t) + t^2 w'(t) \} \hat\ell(t) \ dt } . $$
In particular we considered inverse-probability (IP) -weighted estimators with $\hat\mu=\hat\lambda=0$, regression estimators with $\hat\pi=\infty$, and doubly robust estimators that rely on estimates of all of $(\pi,\lambda,\mu)$. {Note that although we recommend the doubly robust estimators in practice, even these IP and regression estimators are novel and have not been previously proposed, to the best of our knowledge}. For the weight function $w(t)$ we used the density of a $\text{TN}\{0,1,(-1.9,1.9)\}$ truncated normal variable, which roughly matches the marginal distribution of $Z$. To misspecify models for $(\hat\pi,\hat\lambda,\hat\mu)$ we used the \textcite{kang2007demystifying} covariate transformations; to misspecify $\hat\lambda$ we additionally transformed $Z$ to $e^Z$ and used a logit rather than probit model. We used maximum likelihood to estimate $\hat\pi$ and $\hat\lambda$, and a generalized additive model to estimate $\hat\mu$. {Coverage was assessed based on bootstrap confidence intervals, using 100 bootstrap samples.} Results (including bias, standard errors, root mean squared error (RMSE), and coverage) are shown in Figure~\ref{fig:simres}.

The weighting estimator gives large bias unless its nuisance estimator $\hat\pi$ is correctly specified, and similarly the regression estimator gives large bias unless $\hat\lambda$ and $\hat\mu$ are correctly specified; however, the doubly robust estimator gives small bias as long as either $\hat\pi$ or $(\hat\lambda,\hat\mu)$ are correct. In our setup, even under misspecification, the bias for the doubly robust estimator is smaller than that of the weighting or regression estimators. The weighted estimator is least precise, while the regression estimator is most precise, and can outperform the doubly robust estimator in terms of mean squared error when both $\hat\lambda$ and $\hat\mu$ are estimated well, particularly for smaller sample sizes (indicating a bias-variance trade-off). {Coverage roughly coincided with bias: the weighting and regression estimators gave poor coverage unless their nuisance estimators were correctly specified, but the doubly robust estimator gave good coverage as long as either $\hat\pi$ or $(\hat\lambda,\hat\mu)$ was modeled well. Correctly-specified weighting and doubly robust estimators exhibited some conservative behavior (i.e., greater than 95\% coverage) at $n=1000$, and slightly anti-conservative behavior for the larger sample size; but this may be explained by the relatively modest number of bootstrap samples and simulations (which was required to reduce computation time).}

{The regression estimator gave slight bias and relatively poor coverage even for large sample sizes; we hypothesize that this is because this estimator relies on a generalized additive model estimator, which may not be smooth enough to guarantee $\sqrt{n}$ rates and asymptotic linearity. As noted throughout our paper, this is an important motivation for using doubly robust and other influence-function-based estimators, which can attain $\sqrt{n}$ rates and asymptotic linearity even if their nuisance estimators do not. In the Supplementary Materials we present simulation results based on using flexible methods (generalized additive models and random forests) to estimate all of $(\pi,\lambda,\mu)$, not just $\mu$. }

\section{Illustration}

In this section we apply the proposed methodology to estimate the effects on infant mortality of delivery at hospitals with high- versus low-level neonatal intensive care units (NICUs). Following \textcite{lorch2012differential} and others, we define high-level NICUs as those that are designated as level III by the American Academy of Pediatrics, and that deliver at least 50 low birthweight infants on average per year. Level III units have high technical capacity, providing subspecialist teams, advanced imaging, and the ability for sustained mechanical assisted ventilation. On the other hand, level I-II NICUs are only designed to provide basic care to lower-risk infants. The question of whether and how care at high-level NICUs might impact infant mortality is important from both patient and policy perspectives. For example if high-level units can reduce infant mortality, particularly among high-risk infants, then policies that send high-risk infants to high-level NICUs might be worth pursuing. 

\begin{figure}[h!]
\begin{center}
\includegraphics[width=\textwidth]{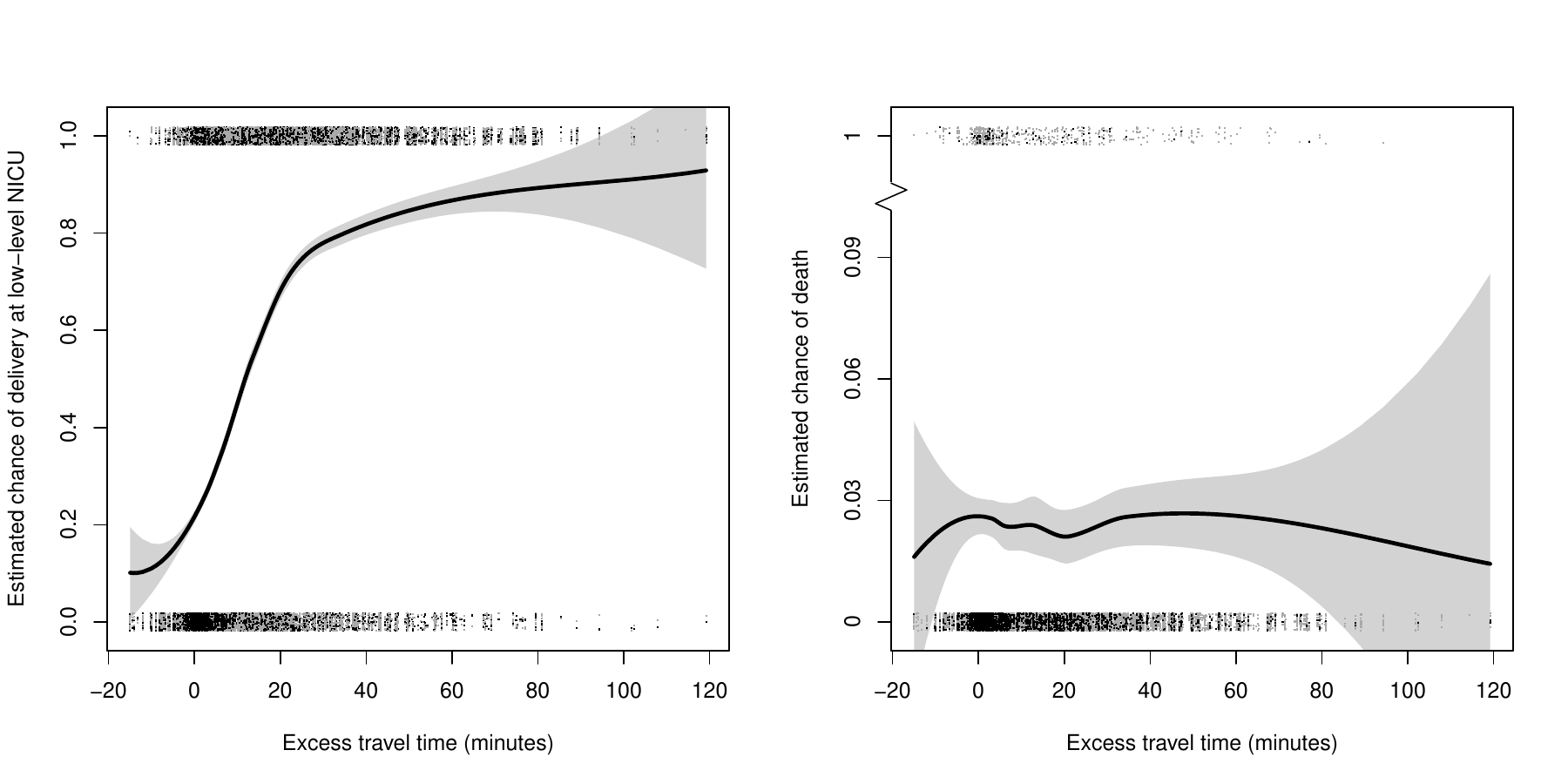}
\end{center}
\caption{Relationship between instrument $Z$ (excess travel time) and treatment $A$ (delivery at low-level unit) on the left, and instrument and outcome $Y$ (infant mortality) on the right. Note that $y$-axis scaling differs in the two plots. \label{fig:data}}
\end{figure}

To assess potential benefits of delivery at hospitals with high-level units, \textcite{lorch2012differential} collected data on all $n=192,078$ premature births in Pennsylvania between 1995 and 2006. Covariate information included data about the infant, such as birthweight and gestational age, as well as about the delivering mother, such as age, race, and measures of socioeconomic status and comorbidities. {A full list of covariate information is given in the Supplementary Materials, and} more details can be found in \textcite{baiocchi2010building} and \textcite{lorch2012differential}. Importantly, the data are missing some detailed clinical information (e.g., comorbidity severity and lab results) that might explain mothers' deliveries at high- versus low-level hospitals; therefore analyses relying on `no unmeasured confounding' assumptions could be suspect. Fortunately,  \textcite{baiocchi2010building} and \textcite{lorch2012differential} identified a potential IV, which is the excess travel time (in minutes) it takes a mother to get to the nearest high- versus low-level NICU. This is a plausible instrument since it affects where mothers deliver (larger values mean mothers have to travel longer to get to high-level units), but it likely does not independently affect infant mortality and is probably not associated with unmeasured confounders that also affect mortality (at least conditional on measured factors like socioeconomic status). More discussion can be found in \textcite{baiocchi2010building} and \textcite{lorch2012differential}. Figure \ref{fig:data} shows loess fits of the unadjusted relationship between instrument and treatment (which is strong), and between  instrument and outcome (which is less strong); the raw data points also indicate the marginal distribution of the instrument. The gray regions denote pointwise 95\% confidence intervals. 

We conducted two sets of analyses based on the methodology proposed in previous sections. First we estimated the LIV curve only conditional on the threshold value (so that $\bV=\emptyset$), and used the proposed cross-validation approach to select among spline models. Second we estimated how effects vary with birthweight and gestational age, which are two important potential effect modifiers. In both analyses it is first necessary to estimate the nuisance functions, which we did using generalized additive models. To estimate the instrument density $\pi$, we used a model previously used by \textcite{kennedy2017nonparametric}, in which the density only depends on covariates through the mean and variance functions but is otherwise flexible. Specifically this model assumes $Z = \pi_1(\bX) + \pi_2(\bX)\epsilon$, where $\epsilon$ satisfies $\E(\epsilon \mid \bX)=0$ and $\E(\epsilon^2 \mid \bX)=1$, the density $f_\epsilon$ of $\epsilon$ is unspecified but smooth, and $(\pi_1,\pi_2)$ follow generalized additive models with identity and log links, respectively. Thus under this model the conditional density of the instrument is given by $\pi(z \mid \bx) = f_\epsilon[\{ z-\pi_1(\bx)\} /\pi_2(\bx) ]$. 

In our first analysis we estimated the LIV curve $\gamma(t)$ using a density-weighted projection based on the marginal density of the instrument, so that $w(t)=\hat{p}(Z=t)$ for $\hat{p}$ a usual kernel density estimator. {A plot of this weight function is given in the Supplementary Materials. The weighted average treatment effects we report should therefore be interpreted as averages of the true LIV curve values, where the averaging weights non-extreme moderate instrument values most highly, according to the weight function plotted in the Supplementary Materials. As noted in Section 3.2, when projecting onto a constant, the resulting weighted average effect also equals the complier average effect in a population where the density of the latent threshold $T$ equals the weight function $w^*$.} We used natural cubic splines for $\gamma_k(t;\bpsi_k)$ with degrees of freedom $k \in \{1,2,3,4\}$ selected via cross-validation with two folds, using the proposed doubly robust pseudo-risk estimate $\hat{R}(\hat\gamma_k)$. 

The linear model with $k=2$ gave the smallest pseudo-risk ($-13.7 \times 10^{-6}$), much smaller than $k=3$ and $k=4$, which led to overfitting ($-7.3 \times 10^{-6}$ and $1.5 \times 10^{-3}$). The risk under $k=2$ was similar to that of the constant effect model with $k=1$ ($-12.6 \times 10^{-6}$) and gave very similar estimates. For example, for the linear model the effect estimates range from 9.0 to 8.9 deaths per 1000 births for excess travel times ranging from 0 to 100, and at level 0.05 we cannot reject the hypothesis that the slope parameter equals zero ($p=0.98$). Table \ref{tab:restab1} gives estimates and 95\% confidence intervals (based on the bootstrap) for three estimators using the constant effect working model $\gamma(t;\bpsi)=\psi$;  the inverse-probability-weighted estimator only relies on estimating the conditional instrument density $\pi$ (i.e., it plugs in sample averages of $A$ and $Y$ for $\hat\lambda$ and $\hat\mu$), the regression-based estimator only relies on estimating the treatment and outcome regressions $(\lambda,\mu)$ (i.e., it plugs in $\infty$ for $\hat\pi$), and the doubly robust estimator is the proposed approach detailed in Section 3.

\begin{table}[h!]
\caption{Risk difference estimates using weighted average of LIV curve (in terms of deaths prevented per 1000 births). \label{tab:restab1}}
\begin{center}
\vspace{-.2in}
\begin{tabular}{l|r}
Method & Est (95\% CI) \\
\hline
Inverse-probability-weighted & -4.8 (-17.2, 7.6) \\
Regression-based & 9.2 (6.3, 12.1)  \\
Doubly robust & 8.9 (5.4, 12.5)  \\
\end{tabular}
\end{center}
\end{table}

The proposed doubly robust estimator indicates a mortality benefit (risk difference) of 8.9 fewer deaths per 1000 births due to high-level NICU care (95\% CI: 5.4, 12.5), among compliers who could be encouraged by travel time to go to a low-level unit. For comparison, this estimate contrasts with the unadjusted risk difference of -18.6 (-20.0, -17.2), which makes high-level NICUs appear to be harming infants, and a doubly robust no-unmeasured-confounding-based estimate of -0.6 (-2.8, 1.6) for the average treatment effect, which does not give any evidence of benefit. Our estimator yields similar results as the two-stage least squares and matching analyses in \textcite{baiocchi2010building}, but targets a different parameter and relies on weaker assumptions. Our regression and doubly robust estimates were similar, and differed somewhat from weighting, indicating that the conditional density model might be misspecified (although the weighted estimator is also imprecise). 

In our second analysis (exploring effect modification by birthweight and gestational age), we projected onto a model in which effects do not vary with the latent threshold (based on the results of our first analysis) but can vary with normal versus low birthweight (2000+ grams versus $<$2000 grams) and early versus very early gestational age (35--37 weeks versus $\leq$34 weeks). Therefore in this analysis we set $\gamma(t,v;\bpsi) = \sum_j \psi_j \one(v=j)$ where $j \in \{1,2,3,4\}$ indexes the four groups. Results are given in Table \ref{tab:restab2}. 

\begin{table}[h!]
\caption{Effect estimates (95\% confidence intervals) by birthweight and gestational age. \label{tab:restab2}}
\begin{center}
\vspace{-.2in}
\begin{tabular}{l|rr}
 & \multicolumn{2}{c}{{Gestational age}} \\
{Birthweight} & $\leq 34$ wks & 35--37 wks \\
\hline
$<$2000 g & 58.5 (52.7, 64.3) & 6.1 (2.2, 10.0) \\
2000+ g & 10.1 (2.7, 17.4) & 2.4 (-0.8, 5.6) \\
\end{tabular}
\end{center}
\end{table}

The largest effect of high-level NICU care was for the highest-risk infants with low birthweight and very early gestational age; in particular, for this group, care at high-level NICUs was estimated to yield 58.5 fewer deaths per 1000 births (95\% CI: 52.7, 64.3). Effects in the other two higher-risk groups were relatively similar, with an estimated 6.1 and 10.1 fewer deaths per 1000 births (and both statistically significantly different from zero). For the lower-risk group with higher birthweight and gestational age, care at high-level units was estimated to yield 2.4 fewer deaths per 1000 births (95\% CI: -0.8, 5.6), and we cannot reject a null hypothesis of zero average effect. Results were similar but less pronounced when we used less extreme cutoffs for birthweight (2500+ grams versus $<$2500 grams) and gestational age ($\leq$ 35 weeks versus 36--37 weeks).

\section{Discussion}

In this paper we developed novel semiparametric theory and estimation procedures for a marginal version of the LIV curve, which represents the effect among local compliers who would be encouraged to take treatment at a given threshold value of the instrument but not below. Importantly, in contrast to available methods for estimating the fully conditional LIV curve, our methods have the following advantages: they do not require parametric assumptions (but can still yield parametric root-n rates of convergence), incorporate information about the instrument mechanism, are doubly robust (i.e., still yield consistent estimates under misspecification of either the instrument or treatment/outcome processes), and allow for estimating varying amounts of effect modification. We described the asymptotic properties of our methods under weak empirical process conditions, and also proposed a doubly robust cross-validation approach for model selection. Finally we used the proposed methods to study the effects of care at high-level NICUs on infant mortality, including how such effects are modified by infants' birthweight and gestational age. 

There are a number of direct opportunities for future work based on this research{, beyond those areas mentioned in the main text (e.g., extensions to deal with positivity violations, loss functions beyond $L_2$, further development and assessment of the sample-splitting estimator, nonparametric estimation of the local IV curve via kernel-smoothed projections, post-selection inference, etc.)}. First, it will be very useful to develop computationally efficient software for implementing the proposed methods for general non-linear working models. The methods are computationally demanding due to the need to calculate multiple derivatives and integrals, especially in cases involving complex effect modification. Second, it might be of interest to determine the efficient choice of the weight function $w(t,\bv)$ for the case where the working model $\gamma(t,\bv;\bpsi)$ is believed to be the true model. A third area of future work is in the application studying the effects of high-level NICU care, where it would be useful to implement more flexible covariate adjustment (e.g., Super Learner) and more complex models (e.g., exploring threshold effects and continuous effect modification). 

\stepcounter{section}
\printbibliography[title={\thesection \ \ \ References}]

\pagebreak

\setcounter{page}{1}

\vspace*{.2in}
\begin{center}
\LARGE \textbf{Supplementary Materials for ``Robust causal inference with continuous  instruments  using the local instrumental variable curve''}
\end{center}

\vspace*{.5in}

\normalsize

\setcounter{section}{0}
\section{Proof of Theorem 1}

First note that  
$$ Z \ind (A^z, Y^z) \mid \bX  \implies  Z \ind (A^z, Y^{zA^z}) \mid \bX \implies  Z \ind (A^z, Y^{A^z}) \mid \bX , $$
where the first implication follows from Assumption 2 (consistency) and the second by Assumption 5 (exclusion restriction). \\

Therefore
\begin{align*}
\E(Y \mid \bX,Z=z ) &= \E(Y^{A^z} \mid \bX,Z=z) = \E(Y^{A^z} \mid \bX) = \E\{A^z (Y^1 -Y^0) \mid \bX \} + \E(Y^0 \mid \bX) 
\end{align*}
where the first equality follows from Assumption 2 (consistency), the second since Assumption 4 (unconfoundedness of $Z$) implies $Z \ind (A^z, Y^{A^z}) \mid \bX$ under Assumptions 2 (consistency) and 5 (exclusion restriction) as shown above, and the third by rearranging. By the same logic we have
$$ \E(A \mid \bX,Z=z ) = \E(A^z \mid \bX,Z=z) = \E(A^z \mid \bX ) . $$
Assumption 3 (positivity) allows us to write conditional expectations given $\bX$ and $Z$. \\

Therefore, combining the above results gives
$$ \E(Y \mid \bX,Z=z+\delta )  - \E(Y \mid \bX,Z=z )  = \E\{(A^{z+\delta}-A^z) (Y^1 -Y^0) \mid \bX\} $$
and
$$ \E(A \mid \bX,Z=z+\delta )  - \E(A \mid \bX,Z=z )  = \E(A^{z+\delta}-A^z \mid \bX) , $$
so that 
\begin{align*}
\E\{ \E(Y \mid \bX&,Z=z+\delta )  - \E(Y \mid \bX,Z=z ) \mid \bV\} = \E\{(A^{z+\delta}-A^z) (Y^1 -Y^0) \mid \bV\} \\
&= \E(Y^1 -Y^0 \mid \bV , A^{z+\delta}>A^z ) P(A^{z+\delta}>A^z \mid \bV) \\
&=  \E(Y^1 -Y^0 \mid \bV , z < T \leq z+\delta) P(z < T \leq z+\delta \mid \bV) 
\end{align*}
and similarly
\begin{align*}
\E\{ \E(A \mid \bX&,Z=z+\delta )  - \E(A \mid \bX,Z=z ) \mid \bV\} = \E(A^{z+\delta}-A^z \mid \bV\} \\
&= P(A^{z+\delta}>A^z \mid \bV) =  P(z < T \leq z+\delta \mid \bV) ,
\end{align*}
where the first equalities follow by iterated expectation, the second by Assumption 1 (monotonicity), which implies $A^{z+\delta}-A^z = \one(A^{z+\delta}>A^z)$, and the third by definition of the latent threshold $T$, i.e., $\{A^{z+\delta}>A^z\} \iff \{ A^{z+\delta}=1, A^z=0 \} \iff \{ T \leq z+\delta, T>z \}$.  \\

Therefore, letting $\gamma(\bv,t)=\E(Y^1-Y^0 \mid T=t, \bV=\bv)$, we have
\begin{align*}
\lim_{\delta \rightarrow 0}  \frac{1}{\delta} &\E\{ \E(Y \mid \bX,Z=t+\delta )  - \E(Y \mid \bX,Z=t ) \mid \bV\} \\
&= \lim_{\delta \rightarrow 0} \frac{1}{\delta} \E(Y^1 -Y^0 \mid \bV , t \leq T \leq t+\delta) P(t \leq T \leq t+\delta \mid \bV)  \\
&= \gamma(t,\bV) \ \lim_{\delta\rightarrow 0} \frac{1}{\delta} \Big\{ P(T \leq t+\delta \mid  \bV) - P(T \leq t \mid \bV) \Big\} \\
&= \gamma(t,\bV) \ p(T =t \mid \bV)
\end{align*}
and similarly
\begin{align*}
\lim_{\delta \rightarrow 0} \frac{1}{\delta} &\E\{ \E(A \mid \bX,Z=t+\delta )  - \E(A \mid \bX,Z=t ) \mid \bV\} = \lim_{\delta \rightarrow 0} \frac{1}{\delta} P(t \leq T \leq t+\delta \mid \bV)  \\
&= \lim_{\delta\rightarrow 0} \frac{1}{\delta} \Big\{ P(T \leq t+\delta \mid \bV) - P(T \leq t \mid \bV) \Big\} \\
&= p(T =t \mid \bV) ,
\end{align*}
where the equalities follow by Assumption 7 (continuity). Specifically, the first and third equalities follow by the fact that $T$ is continuously distributed, with $p(T=t \mid \bV) = \frac{\partial}{\partial t} P(T \leq t \mid \bV)$, and the second follows by the continuity of $\gamma(\bv,t)$ in $t$. \\

Therefore
$$ \gamma(t,\bv) =  \frac{ \frac{\partial}{\partial z} \E\{ \E(Y \mid \bX,Z=z ) \mid \bV=\bv\}  }{ \frac{\partial}{\partial z} \E\{ \E(A \mid \bX,Z=z ) \mid \bV=\bv\} } \bigg|_{z=t} $$
since the denominator is bounded away from zero by Assumption 6 (instrumentation). \\

\section{Proof of Theorem 2}

In this section we use subscripts to index quantities that depend on the distribution $P$; a zero subscript denotes a quantity evaluated at the true distribution $P=P_0$. Thus for example $\E_P$ denotes expectations under $P$ and $\E_0$ denotes expectations under the truth $P=P_0$; similarly $\bpsi_P$ denotes the parameter $\bpsi=\bpsi(P)$ as a map $\bpsi: P \mapsto \R^q$ and $\bpsi_0$ denotes its true value evaluated at $P_0$. \\

In the interest of notational simplicity we make two slight abuses of notation. First, in the main text $\boldsymbol\varphi$ was only proportional to the efficient influence function (i.e., $\boldsymbol\varphi$ did not include the constant matrix scaling factor, which is unnecesary for solving estimating equations), whereas in this appendix we use $\boldsymbol\varphi$ to denote the full efficient influence function (including the constant matrix scaling factor). Second, the functions $\bg_j(t,\bv;\bpsi)$ (for $j=1,2$) in the main text are denoted by $\bg'_j(t,\bv;\bpsi)$ in this appendix. \\

We will show that $\boldsymbol\varphi(\bO;\bpsi_P,\boldsymbol\eta_P)=\boldsymbol\varphi_P(\bO)$ is the efficient influence function by showing that it is the canonical gradient of the pathwise derivative of $\bpsi_P$, i.e., that $\boldsymbol\varphi_P$ satisfies
$$ \frac{\partial \bpsi_\epsilon}{\partial \epsilon} \Bigm|_{\epsilon=0}  = \E_0\{ \boldsymbol\varphi_0(\bO) s_0(\bO) \}  $$
where $\bpsi_\epsilon = \bpsi(P_\epsilon)$ denotes the parameter $\bpsi$ evaluated at any regular parametric submodel $\{P_\epsilon : \epsilon\}$ passing through $P_0$ at $\epsilon=0$, and $s_\epsilon(\bo_1 \mid \bo_2) = \frac{\partial}{\partial\epsilon^*} \log dP_\epsilon^*(\bo_1 \mid \bo_2)  |_{\epsilon^*=\epsilon}$ denotes the parametric submodel score for any partition $(\bO_1, \bO_2) \subseteq \bO$. \\

By definition we have
\begin{align*}
\bpsi_P &= \argmin_{\bpsi \in \R^q} \int_{\mathcal{V}} \int_{\mathcal{T}} w(t,\bv) \Big\{ \gamma_P(t,\bv)-\gamma(t,\bv;\bpsi) \Big\}^2 p(T=t \mid \bv) \ dt \ dP(\bv) 
\end{align*}
and thus
$$ \int_{\mathcal{V}} \int_{\mathcal{T}} \frac{\partial \gamma(t,\bv;\bpsi)}{\partial \bpsi}\Bigm|_{\bpsi=\bpsi_P} w(t,\bv) \Big\{ \gamma_P(t,\bv)-\gamma(t,\bv;\bpsi_P) \Big\} p(T=t \mid \bv) \ dt \ dP(\bv) = 0 . $$ 

\medskip

Letting $m_P(z,\bv)=\E_P\{\E_P(Y \mid \bX, Z=z) \mid \bV=\bv\}$ and $m_P'(t,\bv)=\frac{\partial}{\partial z} m_P(z,\bv)|_{z=t}$, and similarly $\ell_P(z,\bv)=\E_P\{\E_P(A \mid \bX, Z=z) \mid \bV=\bv\}$ and $\ell_P'(t,\bv)=\frac{\partial}{\partial z} \ell_P(z,\bv)|_{z=t}$, then under the identifying assumptions in the main text we have
$$ \gamma_P(t,\bv) = \frac{m_P'(t,\bv)}{\ell_P'(t,\bv)} \ \text{ and } \ p(t \mid \bv) = \ell_P'(t,\bv) . $$

\medskip

Therefore the restriction above is equivalent to
\begin{align*}
0 &= \int_{\mathcal{V}} \int_{\mathcal{T}} \frac{\partial \gamma(t,\bv;\bpsi)}{\partial \bpsi}\Bigm|_{\bpsi=\bpsi_P} w(t,\bv) \Big\{ m_P'(t,\bv)  - \gamma(t,\bv;\bpsi_P) \ell_P'(t,\bv) \Big\} \ dt \ dP(\bv) \\
&= \int_{\mathcal{V}} \int_{\mathcal{T}} \Big\{ \bg_2(t,\bv;\bpsi_P) \ m_P'(t,\bv)  - \bg_1(t,\bv;\bpsi_P) \ \ell_P'(t,\bv) \Big\} \ dt \ dP(\bv)
\end{align*}
where $\bg_1$ and $\bg_2$ are $q$-vectors (with known functional form not depending on $P$) defined as
$$ \bg_1(t,\bv;\bpsi) = \bg_2(t,\bv;\bpsi) \gamma(t,\bv;\bpsi) \ \text{ and }  \ \bg_2(t,\bv;\bpsi) =  \frac{\partial \gamma(t,\bv;\bpsi^*)}{\partial \bpsi^*}\Bigm|_{\bpsi^*=\bpsi} w(t,\bv)  . $$

And since the weight satisfies $w(t,\bv)=0$ for $t \notin \text{int}(\mathcal{T})$,  integration by parts gives
$$ \int_{\mathcal{V}} \int_{\mathcal{T}} \Big\{ \bg'_1(t,\bv;\bpsi_P) \ \ell_P(t,\bv)  - \bg'_2(t,\bv;\bpsi_P) \ m_P(t,\bv)  \Big\} \ dt \ dP(\bv) = 0 , $$
where $\bg'_j(t,\bv;\bpsi)=\partial \bg_j(z,\bv;\bpsi)/\partial z |_{z=t}$. \\

Evaluating the above at $P=P_\epsilon$ gives 
$$ \int_{\mathcal{V}} \int_{\mathcal{T}} \Big\{ \bg'_1(t,\bv;\bpsi_\epsilon) \ell_\epsilon(t,\bv)  - \bg'_2(t,\bv;\bpsi_\epsilon) m_\epsilon(t,\bv)  \Big\} \ dt \ dP_\epsilon(\bv) = 0 , $$
and differentiating with respect to $\epsilon$ and evaluating at the truth $\epsilon=0$ (using the chain rule) gives 
\begin{align*}
0 &= \int_{\mathcal{V}} \int_{\mathcal{T}} \bigg\{ \frac{\partial \bg'_1(t,\bv;\bpsi)}{\partial \bpsi} \Bigm|_{\bpsi=\bpsi_0} \frac{\partial \bpsi_\epsilon}{\partial \epsilon}\Bigm|_{\epsilon=0} \ell_0(t,\bv) + \bg'_1(t,\bv;\bpsi_0) \frac{\partial \ell_\epsilon(t,\bv)}{\partial \epsilon}\Bigm|_{\epsilon=0} \\
& \hspace{.5in} - \frac{\partial \bg'_2(t,\bv;\bpsi)}{\partial \bpsi} \Bigm|_{\bpsi=\bpsi_0} \frac{\partial \bpsi_\epsilon}{\partial \epsilon}\Bigm|_{\epsilon=0} m_0(t,\bv) - \bg'_2(t,\bv;\bpsi_0) \frac{\partial m_\epsilon(t,\bv)}{\partial \epsilon}\Bigm|_{\epsilon=0}  \bigg\} dt \ dP_0(\bv) \\
& \hspace{.25in} + \int_{\mathcal{V}} \int_{\mathcal{T}} \Big\{ \bg'_1(t,\bv;\bpsi_0) \ell_0(t,\bv)  - \bg'_2(t,\bv;\bpsi_0) m_0(t,\bv)  \Big\} s_0(\bv) \ dt \ dP_0(\bv) .
\end{align*}

\medskip

Rearranging, this implies that
\begin{align*}
\frac{\partial \bpsi_\epsilon}{\partial \epsilon}\Bigm|_{\epsilon=0} &= \mathbf{C}_0^{-1} \int_{\mathcal{V}} \int_{\mathcal{T}} \bigg[ \bg'_1(t,\bv;\bpsi_0) \left\{ \frac{\partial \ell_\epsilon(t,\bv)}{\partial \epsilon}\Bigm|_{\epsilon=0}  + \ell_0(t,\bv) s_0(\bv) \right\} \\
& \hspace{.5in} - \bg'_2(t,\bv;\bpsi_0) \left\{ \frac{\partial m_\epsilon(t,\bv)}{\partial \epsilon}\Bigm|_{\epsilon=0} + m_0(t,\bv) s_0(\bv) \right\} \bigg] dt \ dP_0(\bv)
\end{align*}
with
\begin{align*}
\mathbf{C}_P = & -\int_{\mathcal{V}} \int_{\mathcal{T}} \left\{ \frac{ \partial \bg'_1(t,\bv;\bpsi) }{\partial \bpsi}\Bigm|_{\bpsi=\bpsi_P} \ell_P(t,\bv) - \frac{ \partial \bg'_2(t,\bv;\bpsi) }{\partial \bpsi}\Bigm|_{\bpsi=\bpsi_P} m_P(t,\bv) \right\} dt \ dP(\bv) ,
\end{align*}
and
\begin{align*}
\frac{\partial \ell_\epsilon(z,\bv)}{\partial \epsilon}\Bigm|_{\epsilon=0} &= \frac{\partial}{\partial \epsilon} \E_\epsilon \{ \E_\epsilon (A \mid \bX, Z=z) \mid \bV=\bv\} |_{\epsilon=0} = \frac{\partial}{\partial \epsilon} \int_\mathcal{W} \sum_{a \in \{0,1\}} a \ p_\epsilon(a \mid \bx, z) \  dP_\epsilon(\bw \mid \bv) \Big|_{\epsilon=0} \\
&= \int_\mathcal{W} \sum_{a \in \{0,1\}} a \ \Big\{ s_0(a \mid \bx, z) + s_0(\bw \mid \bv) \Big\} \ p_0(a \mid \bx, z)  \ dP_0(\bw \mid \bv) \\
&= \E_0\Big(\E_0\Big[ A \{ s_0(A \mid \bX, Z) + s_0(\bW \mid \bV) \} \Bigm| \bX, Z=z \Big] \Bigm| \bV=\bv\Big) ,
\end{align*}
and by the same logic
$$ \frac{\partial m_\epsilon(z,\bv)}{\partial \epsilon}\Bigm|_{\epsilon=0} = \E_0\Big(\E_0\Big[ Y \{ s_0(Y \mid \bX, Z) + s_0(\bW \mid \bV) \} \Bigm| \bX, Z=z \Big] \Bigm| \bV=\bv\Big) . $$

Now we turn to $\E_0\{ \boldsymbol\varphi_0(\bO) s_0(\bO) \}$. The putative efficient influence function $\boldsymbol\varphi_P$ from the main text is given by
\begin{align*}
\boldsymbol\varphi_P(\bO) &= \mathbf{C}_P^{-1}  \bigg[ \bg'_1(Z,\bV;\bpsi_P) \bigg\{ \frac{A - \E_P(A \mid \bX,Z)}{p(Z \mid \bX)} \bigg\} - \bg'_2(Z,\bV;\bpsi_P) \bigg\{ \frac{Y - \E_P(Y \mid \bX,Z)}{p(Z \mid \bX)} \bigg\} \\
& \hspace{.5in} + \int_{\mathcal{T}} \Big\{ \bg'_1(t,\bV;\bpsi_P) \E_P(A \mid \bX,Z=t) - \bg'_2(t,\bV;\bpsi_P) \E_P(Y \mid \bX, Z=t) \Big\} \ dt \bigg] 
\end{align*}
(note the inclusion of the scaling matrix $\mathbf{C}_P^{-1}$), and $s_0(\bO)$ is the parametric submodel score, which can be decomposed as
$$ s_0(\bO) = s_0(Y, A \mid \bX, Z) + s_0(Z \mid \bX) + s_0(\bW \mid \bV) + s_0(\bV) . $$ 

\medskip

Therefore
\begin{align*}
\mathbf{C}_0 &\E_0\Big[ \boldsymbol\varphi_0(\bO) \{ s_0(Y, A \mid \bX, Z) + s_0(Z \mid \bX) + s_0(\bW \mid \bV) + s_0(\bV) \} \Big] \\
&= \E_0 \bigg[ \bg'_1(Z,\bV;\bpsi_0) \bigg\{ \frac{A s_0(A \mid \bX, Z) }{p_0(Z \mid \bX)} \bigg\}  - \bg'_2(Z,\bV;\bpsi_0) \bigg\{ \frac{Y s_0(Y \mid \bX, Z) }{p_0(Z \mid \bX)} \bigg\} \\
& \hspace{.5in} + \int_{\mathcal{T}} \Big\{ \bg'_1(t,\bV;\bpsi_0) \E_0(A \mid \bX,Z=t) - \bg'_2(t,\bV;\bpsi_0) \E_0(Y \mid \bX, Z=t) \Big\} \ dt \\
& \hspace{1in} \times \Big\{ s_0(\bW \mid \bV) + s_0(\bV) \Big\} \bigg] \\
&=\int_\mathcal{V} \int_{\mathcal{T}} \bigg( \bg'_1(t,\bv;\bpsi_0) \E_0\Big( \E_0\Big[ A \{ s_0(A \mid \bX, Z) + s_0(\bW \mid \bV) \} \Bigm| \bX, Z=t \Big] \Bigm| \bV=\bv \Big) \\
& \hspace{.5in} - \bg'_2(t,\bv;\bpsi_0) \E_0\Big( \E_0\Big[ Y \{ s_0(Y \mid \bX, Z) + s_0(\bW \mid \bV) \} \Bigm| \bX, Z=t \Big] \Bigm| \bV=\bv \Big) \\
& \hspace{.5in} + \Big\{ \bg'_1(t,\bv;\bpsi_0) \E_0(A \mid \bX,Z=t) - \bg'_2(t,\bv;\bpsi_0) \E_0(Y \mid \bX, Z=t) \Big\} s_0(\bv) \bigg) \ dt \ dP_0(\bv) \\
&= \int_{\mathcal{V}} \int_{\mathcal{T}} \bigg[ \bg'_1(t,\bv;\bpsi_0) \bigg\{ \frac{\partial \ell_\epsilon(z,\bv)}{\partial \epsilon}\Bigm|_{\epsilon=0}  + \ell_0(t,\bv) s_0(\bv) \bigg\} \\
& \hspace{.5in} - \bg'_2(t,\bv;\bpsi_0) \bigg\{ \frac{\partial m_\epsilon(t,\bv)}{\partial \epsilon}\Bigm|_{\epsilon=0} + m_0(t,\bv) s_0(\bv) \bigg\} \bigg] dt \ dP_0(\bv) = \mathbf{C}_0 \frac{\partial \bpsi_\epsilon}{\partial \epsilon}\Bigm|_{\epsilon=0}
\end{align*}
where the first equality follows by iterated expectation and the fact that $\E_0\{ s_0(\bO_1 \mid \bO_2) \mid \bO_2\}=0$ for any $(\bO_1,\bO_2) \subseteq \bO$, the second follows by iterated expectation, the third follows by iterated expectation and by definition of $\ell_0$ and $m_0$ (along with the earlier results for their derivatives with respect to $\epsilon$), and the fourth follows by the expression derived earlier for ${\partial \bpsi_\epsilon} / {\partial \epsilon}|_{\epsilon=0}$. \\

Therefore, as long as $\mathbf{C}_0$ is invertible, we have ${\partial \bpsi_\epsilon} / {\partial \epsilon}|_{\epsilon=0} = \E_0\{ \boldsymbol\varphi_0(\bO) s_0(\bO) \}$ and thus  $\boldsymbol\varphi_P(\bO)$ is the efficient influence function.

\section{Double robustness of efficient influence function $\boldsymbol\varphi$}

Here we will show that $\E\{ \boldsymbol\varphi(\bO; \bpsi,\overline\pi,\overline\lambda,\overline\mu)\}=0$ as long as either 
$$\overline\pi=\pi_0 \ \text{ or } \ (\overline\lambda,\overline\mu)=(\lambda_0,\mu_0). $$ 
In this section expectations $\E=\E_0$ and parameters $\bpsi=\bpsi_0$ are evaluated under $P_0$, but we drop the subscript for notational convenience. \\

First note that
\begin{align*}
\mathbf{C}_0 & \E\{ \boldsymbol\varphi(\bO; \bpsi,\overline\pi,\overline\lambda,\overline\mu)\} = \bigg[ \bg'_1(Z,\bV;\bpsi) \bigg\{ \frac{A - \overline\lambda(\bX,Z)}{\overline\pi(Z \mid \bX)} \bigg\} - \bg'_2(Z,\bV;\bpsi) \bigg\{ \frac{Y - \overline\mu(\bX,Z)}{\overline\pi(Z \mid \bX)} \bigg\} \\
& \hspace{1.7in} + \int_{\mathcal{T}} \Big\{ \bg'_1(t,\bV;\bpsi) \overline\lambda(\bX,t) - \bg'_2(t,\bV;\bpsi) \overline\mu(\bX, t) \Big\} \ dt \bigg] \\
&= \E\bigg[ \bg'_1(Z,\bV;\bpsi) \bigg\{ \frac{\lambda_0(\bX,Z)- \overline\lambda(\bX,Z)}{\overline\pi(Z \mid \bX)} \bigg\} - \bg'_2(Z,\bV;\bpsi) \bigg\{ \frac{\mu_0(\bX,Z) - \overline\mu(\bX,Z)}{\overline\pi(Z \mid \bX)} \bigg\} \\
& \hspace{.5in} + \int_{\mathcal{T}} \Big\{ \bg'_1(t,\bV;\bpsi) \overline\lambda(\bX,t) - \bg'_2(t,\bV;\bpsi) \overline\mu(\bX, t) \Big\} \ dt \bigg] \\
&= \E\int_{\mathcal{Z}} \bigg[ \bg'_1(t,\bV;\bpsi) \Big\{ \lambda_0(\bX,t)- \overline\lambda(\bX,t) \Big\}  - \bg'_2(t,\bV;\bpsi) \Big\{ \mu_0(\bX,t) - \overline\mu(\bX,t) \Big\} \bigg] \frac{\pi_0(t \mid \bX)}{\overline\pi(t \mid \bX)} \ dt \\
& \hspace{.5in} + \int_{\mathcal{T}} \Big\{ \bg'_1(t,\bV;\bpsi) \overline\lambda(\bX,t) - \bg'_2(t,\bV;\bpsi) \overline\mu(\bX, t) \Big\} \ dt \bigg] \\
&= \E\int_{\mathcal{T}} \bigg[ \bg'_1(t,\bV;\bpsi) \Big\{ \lambda_0(\bX,t)- \overline\lambda(\bX,t) \Big\}  - \bg'_2(t,\bV;\bpsi) \Big\{ \mu_0(\bX,t) - \overline\mu(\bX,t) \Big\} \bigg] \bigg\{ \frac{\pi_0(t \mid \bX)}{\overline\pi(t \mid \bX)} - 1\bigg\} dt  \\
& \hspace{.5in} + \int_{\mathcal{T}} \Big\{ \bg'_1(t,\bV;\bpsi) \lambda_0(\bX,t) - \bg'_2(t,\bV;\bpsi) \mu_0(\bX, t) \Big\} \ dt \bigg]
\end{align*}
where the first equality is true by definition, the second and third holds by iterated expectation given $(\bX,Z)$ and $\bX$, respectively, and the last follows after rearranging and since $\bg_1'=\bg_2'=0$ for $t \notin \text{int}(\mathcal{T})$. \\

Therefore if  $\overline\pi=\pi_0$ or $(\overline\lambda,\overline\mu)=(\lambda_0,\mu_0)$ then $\mathbf{C}_0 \E\{ \boldsymbol\varphi(\bO; \bpsi,\overline\pi,\overline\lambda,\overline\mu)\}$ equals
\begin{align*} 
\int_{\mathcal{T}} \E&\Big\{ \bg'_1(t,\bV;\bpsi) \lambda_0(\bX,t) - \bg'_2(t,\bV;\bpsi) \mu_0(\bX, t) \Big\} \ dt  \\
&= \int_{\mathcal{V}} \int_{\mathcal{T}} \bg'_1(t,\bv;\bpsi) \ \ell_0(t),\bv) - \bg'_2(t,\bv;\bpsi) \ m_0(t,\bv) \Big\} \ dt \ dP(\bv) = 0
\end{align*}
where the first equality follows by iterated expectation given $\bV$ (and by the definitions of $\ell$ and $m$ from the previous section), and the second follows from the restriction given in the previous section (after using integration by parts).

\section{Sample-splitting estimator}

For simplicity, the estimator presented in the main text solves an estimating equation that depends on nuisance estimates $\boldsymbol{\hat\eta}=(\hat\pi,\hat\lambda,\hat\mu)$ constructed from the entire sample. The asymptotic normality of such estimators requires empirical process (e.g., Donsker) conditions that restrict the complexity of the nuisance functions and their estimators. Unfortunately, these conditions may not be satisfied for very adaptive machine learning methods, such as random forests; however, such conditions are not necessary for estimators that use sample splitting to separate nuisance function estimation from estimating equation evaluation. We present such an estimator here, following \textcite{robins2008higher, zheng2010asymptotic, chernozhukov2016double}. \\

Sample splitting has a long history (cf.\ Bickel (1982), Schick (1986), van der Vaart (1998)), though the idea of explicitly combining with machine learning came more recently. \textcite{robins2008higher} (page 379) used sample splitting to avoid Donsker conditions for functional estimation in nonparametric models, although they did not use cross-fitting. The PhD thesis of Ayygari (2010) (subsequently published as Robins et al.\ (2013)) used cross-fitting in semiparametric efficient estimation with cross-validation-based selection of nuisance estimators. \textcite{zheng2010asymptotic} gave a general cross-fitting version of targeted maximum likelihood estimation to avoid Donsker conditions. Belloni et al.\ (2010) used cross-fitting to relax sparsity requirements in a high-dimensional linear instrumental variable setting. Cross-fitting was also studied more recently by \textcite{chernozhukov2016double}. \\

We rely on the notation presented in Section 3.4 of the main text. Again let $\bS=(S_1,...,S_n)$ denote a random variable independent of the sample that splits the data into training ($S_i=0$) and test ($S_i=1$) sets. Here we suppose the split variable $\bS$ takes $v$ different values $\{ \bS_1,...,\bS_v\}$, each with equal probability $1/v$, where $\sum_i S_{iv}=n/v$ for all $v$ and $\sum_v S_{iv}=1$ for all $i$, so that test sets are all of size $n/v$ and each unit is only used in one test set. Define $\Pb_\bs^0$ and $\Pb_\bs^1$ as the sub-empirical distributions for the training data $\{i: S_i=0\}$ and test data $\{i: S_i=1\}$, respectively, for a given split $\bS=\bs$. Therefore, for example, $\boldsymbol{\hat\eta}(\Pb_\bs^0)$ denotes the nuisance function estimates based only on the training set data. Then the proposed sample-splitting version of our estimator is given by the solution to
$$ \E_\bS  \Pb_\bS^1[ \boldsymbol\varphi\{\bO; \boldsymbol\psi, \boldsymbol{\hat\eta}(\Pb_\bS^0) \} ] = \E_\bS  \int \boldsymbol\varphi\{\bo; \boldsymbol\psi, \boldsymbol{\hat\eta}(\Pb_\bS^0) \} \ d\Pb_\bS^1(\bo) = \mathbf{0} ,  $$
with $\boldsymbol\varphi$ as defined in the main text.

\vspace{-.1in}

\section{Proof of Theorem 3}

Since $\hat\bpsi$ is a Z-estimator, Theorem 3 follows directly from Theorem 5.31 of van der Vaart (2000), together with the fact that $\mathbf{C}_0 \Pb \{ \boldsymbol\varphi(\bO; \bpsi, \hat{\boldsymbol\eta}) \} = \mathbf{C}_0 \int \boldsymbol\varphi(\bo; \bpsi, \hat{\boldsymbol\eta}) \ dP(\bo)$ equals 
\begin{align*}
\Pb\int_{\mathcal{T}} & \bigg[ \bg'_1(t,\bV;\bpsi) \Big\{ \lambda_0(\bX,t)- \hat\lambda(\bX,t) \Big\}  - \bg'_2(t,\bV;\bpsi) \Big\{ \mu_0(\bX,t) - \hat\mu(\bX,t) \Big\} \bigg] \bigg\{ \frac{\pi_0(t \mid \bX)}{\hat\pi(t \mid \bX)} - 1\bigg\} dt \\
&= \Pb\left[ \Big\{ \bg'_1 ( \lambda_0 - \hat\lambda)  - \bg'_2 ( \mu_0 - \hat\mu ) \Big\} ( \pi_0 - \hat\pi ) \Big/ \hat\pi \pi_0 \right] \leq C || \bg'_1 ( \lambda_0 - \hat\lambda)  - \bg'_2 ( \mu_0 - \hat\mu )  ||_2  || \pi_0 - \hat\pi ||_2 \\
&= O_p\left\{ \Big(  || \lambda_0 - \hat\lambda ||_2 + ||  \mu_0 - \hat\mu  ||_2 \Big) || \pi_0 - \hat\pi ||_2 \right\}
\end{align*}
where the first line follows by iterated expectation (first given $\bX$ and $Z$, then given $\bX$), the second by multiplying and dividing by $\pi(t \mid \bX)$, the inequality follows by Cauchy-Schwarz (i.e., $\Pb(fg) \leq ||f||_2 ||g||_2$) and boundedness of $1/\hat\pi \pi_0$ (from the Donsker condition and positivity), and the last line by the triangle inequality and boundedness of $\bg'_1$ and $\bg'_2$.

\section{Avoiding Donsker conditions with sample-splitting}

For the asymptotic properties of the sample-splitting estimator presented in Section 4 of the Supplementary Materials, we refer to \textcite{chernozhukov2016double}, who give a thorough analysis of the conditions required for asymptotic linearity.  \\

Conditions (i)--(v) of their Assumption 2.1 are standard regularity and identifiability conditions in Z-estimator problems, which can be verified for sufficiently smooth working models $\gamma(t,\bv;\bpsi)$ and estimating functions with sufficiently well-separated solutions in expectation. Conditions (i)--(vi) of their Assumption 2.2 will be satisfied under the rate conditions given in our Assumption 4 of Theorem 3 in the main text, along with sufficient smoothness of the working models $\gamma(t,\bv;\bpsi)$. Notably, only rate conditions are required for the nuisance estimators $\boldsymbol{\hat\eta}$, and not Donsker or other regularity conditions. \textcite{zheng2010asymptotic} give similar results for sample-splitting-based targeted maximum likelihood estimators, relying on higher-level conditions.

\section{Proof of Theorem 4 \& double robustness of loss $L$}

After replacing $\bg_j$ with $f_j$, these proofs follow precisely the same logic as the proofs given in previous sections of Theorem 2 and of double robustness of the efficient influence function $\boldsymbol\varphi$, respectively, and so are omitted. 

\section{Additional simulation results}

In the simulations in the main text, we used correct models for the nuisance functions (parametric models for $(\pi,\lambda)$ and a generalized additive model for $\mu$), with misspecification coming from covariate transformations. In this section we present results using generalized additive models and random forests for all nuisance estimators, since it is of interest to explore the performance of our approach when all nuisance functions are estimated without relying on parametric models. The nuisance functions $(\lambda,\mu)$ are just regression functions so we simply use generalized additive models and random forests as usual there; since $\pi$ is a density we first use generalized additive models or random forests to estimate the conditional expectation $\E(Z \mid \bX)$, and then use a kernel density estimator to estimate the density depending on the sign of the regression function. Note that these approaches are not necessarily correctly specified (except for estimating $\mu$ with a generalized additive model, which actually holds). As in the main text, we compare performance using both the ``true'' covariates $\bX$ and the ``misspecified'' covariates based on the Kang \& Schafer (2007) transformations. \\

Results are given in Figure 1 below. Since the models were not necessarily correctly specified, the findings are harder to generalize. The inverse-probability-weighted estimator gave some bias in all settings, which suggests that the density estimation procedure could be improved. The regression estimator tended to give small bias when it relied on the true covariates for the $\mu$ estimator, as might be expected. The doubly robust estimator gave smallest bias when it relied on the true covariates for the $\pi$ estimator; it is unclear why there was bias otherwise. \\

The findings were slightly more consistent for RMSE, especially when using random forests. However, for both generalized additive model and random forests approaches, the doubly robust estimator gave smallest average RMSE across simulation settings, with the inverse-probability-weighted estimator yielding highest average RMSE, and the regression estimator somewhere in between.

\begin{figure}[h!]
\begin{center}
\includegraphics[width=.68\textwidth]{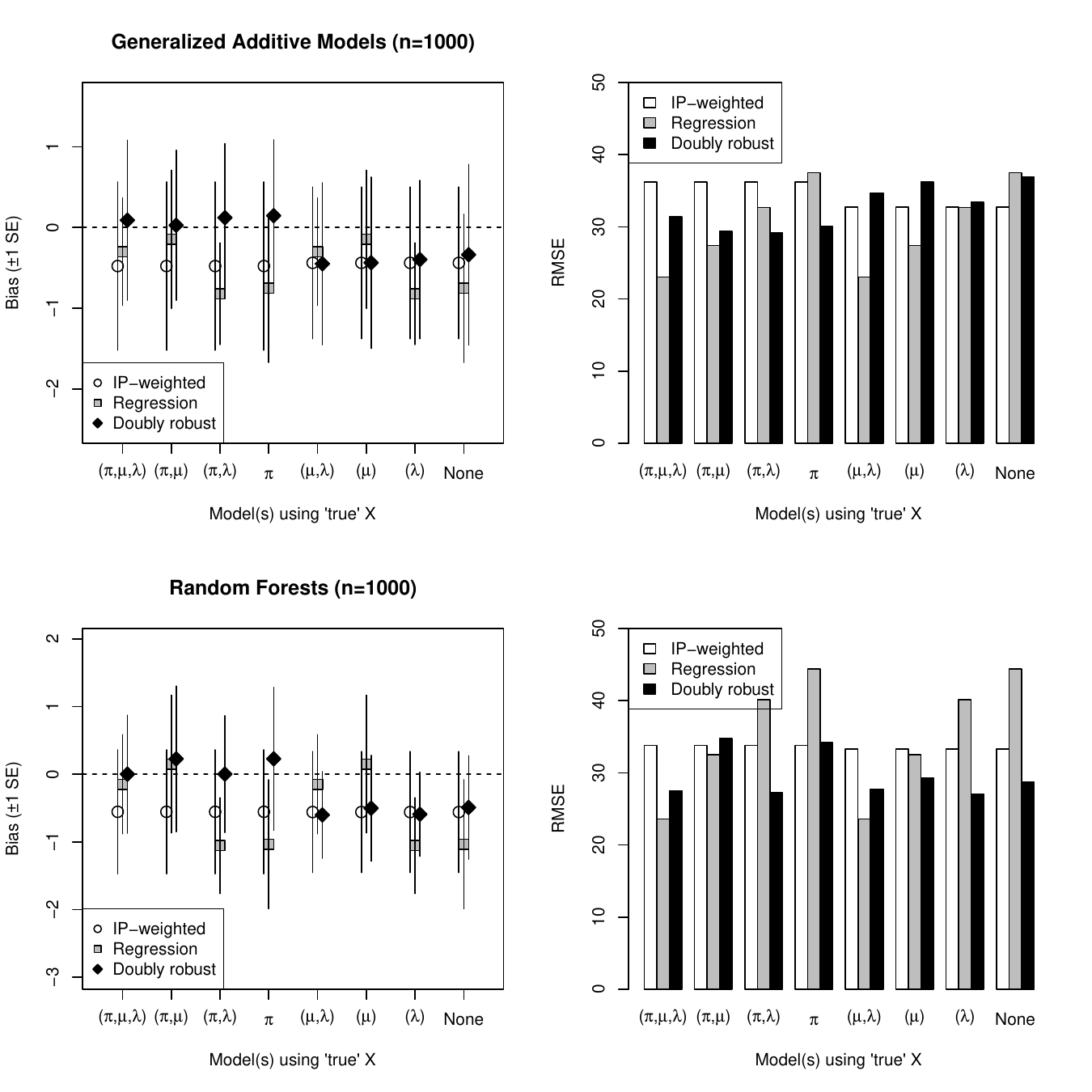}
\end{center}
\caption{Simulation results across 500 simulations: bias (with standard error), RMSE (scaled by $\sqrt{n}$), and bootstrap coverage (based on 100 bootstrap samples). \label{fig:tab}}
\end{figure}

\section{Details about empirical example}

\subsection{List of adjustment covariates}

The data analysis in Section 5 of the main text relied on the following 16 adjustment covariates, as discussed in more detail by Baiocchi et al.\ (2010) and Lorch et al.\ (2012). \\

Information about the zip code in which the mother lives (6):
\begin{itemize}
\item median income, percentage below poverty, median home value, percent with high school degree, percent with college degree, percent who rent versus own home.
\end{itemize}

Information about the mother (8):
\begin{itemize}
\item age, diabetes status, month prenatal care was started, number of times previously given birth, whether multiple deliveries, education level (8th grade or less, some high school, high school graduate, some college, college graduate, or more than college), mother's race (White, Black, Asian/Pacific Islander, or other), insurance type (fee for service, HMO, federal/state, other, or uninsured).
\end{itemize}

Information about the infant (2):
\begin{itemize}
\item birthweight, gestational age.
\end{itemize}

\subsection{Plot of estimated density of $Z$ (i.e., weight function $w$)}

\vspace*{-.3in}
\begin{figure}[h!]
\begin{center}
\includegraphics[height=3.45in]{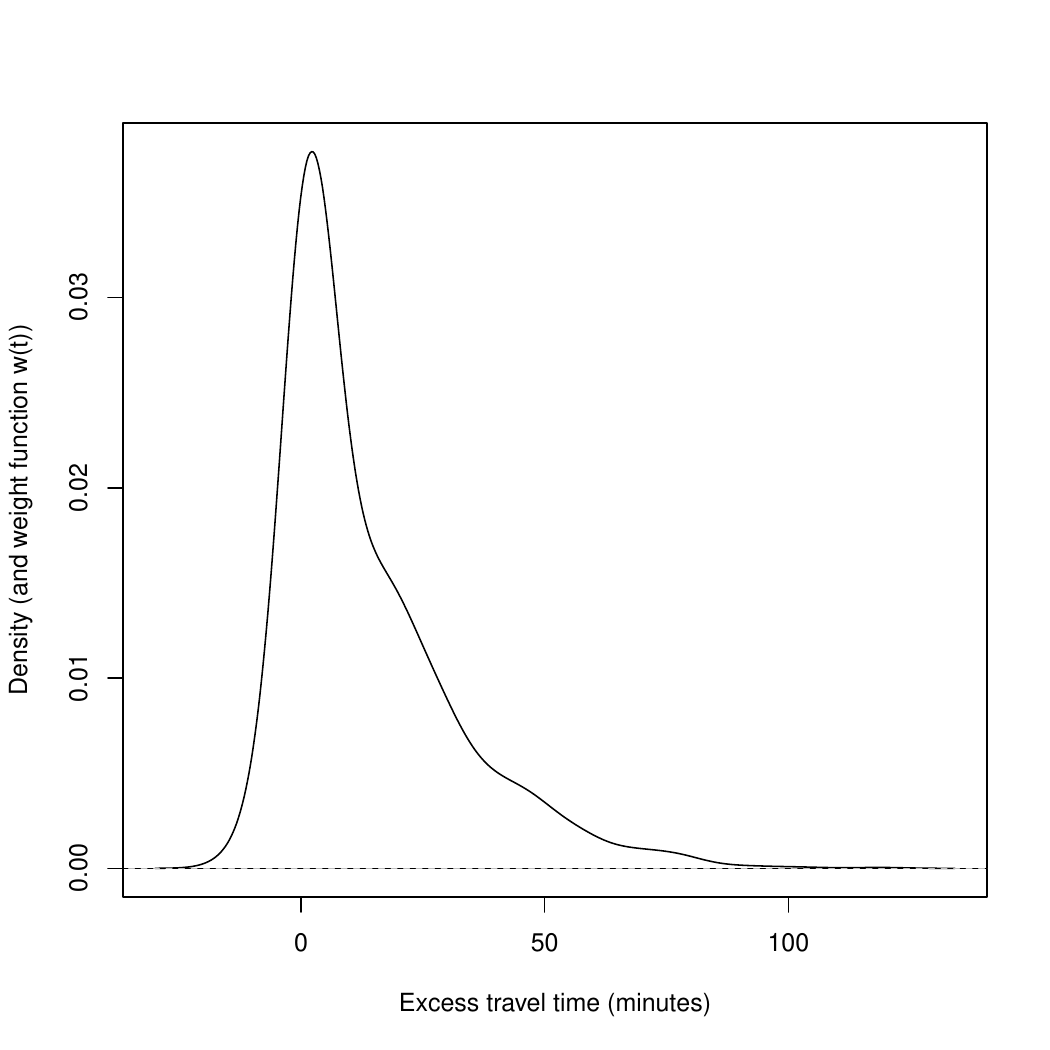}
\end{center}
\caption{Kernel estimate of the marginal density $p(z=t)$ of the instrument $Z$, also used as the weight function $w(t)=\hat{p}(z=t)$ in the analysis in Section 5 of the main text.}
\end{figure}

\end{document}